\documentclass{aa}  

\usepackage{natbib}
\usepackage{graphicx}
\usepackage{txfonts}

\usepackage[dvipsnames]{xcolor}
\usepackage{tikz}
\usetikzlibrary{bayesnet}

\usepackage{diagbox}
\usepackage{subcaption}

\usepackage{comment}
\usepackage{multicol}

\newcommand{\de}{\text{d}}

\newcommand{\kmsec}{\text{km}\,\text{s}^{-1}}

\newcommand{\kpc}{\text{kpc}}
\newcommand{\pc}{\text{pc}} 
\newcommand{\yr}{\text{yr}}
\newcommand{\Myr}{\text{Myr}}

\newcommand{\mas}{\text{mas}}

\newcommand{\magn}{\text{mag}}

\usepackage{hyperref}
\begin{document}

   \title{Mapping Milky Way disk perturbations in stellar number density and vertical velocity using Gaia DR3}
   \titlerunning{Mapping Milky Way disk perturbations}

   \author{A. Widmark
          \inst{1}
          \and
          L. M. Widrow
          \inst{2}
          \and
          A. Naik
          \inst{3}
          }

   \institute{Dark Cosmology Centre, Niels Bohr Institute, University of Copenhagen, Jagtvej 128, 2200 Copenhagen N, Denmark\\
   \email{axel.widmark@nbi.ku.dk}
   \and
   Department of Physics, Engineering Physics, and Astronomy, Queen's University, Kingston K7L 3X5, Canada
   \and
   School of Physics \& Astronomy, University of Nottingham, University Park, Nottingham NG7~2RD, United Kingdom
    }

   \date{Received Month XX, XXXX; accepted Month XX, XXXX}

 
  \abstract{
We have mapped the number density and mean vertical velocity of the Milky Way's stellar disk out to roughly two kiloparsecs from the Sun using \emph{Gaia} Data Release 3 (DR3) and complementary photo-astrometric distance information from StarHorse. For the number counts, we carefully masked spatial regions that are compromised by open clusters, great distances, or dust extinction and used Gaussian processes to arrive at a smooth, non-parametric estimate for the underlying number density field. We find that the number density and velocity fields depart significantly from an axisymmetric and mirror-symmetric model. These departures, which include projections of the {\emph Gaia} phase-space spiral, signal the presence of local disturbances in the disk. We identify two features that are present in both stellar number density and mean vertical velocity. One of these features appears to be associated with the Local Spiral Arm. It is most prominent at small heights and is largely symmetric across the mid-plane of the disk. The density and velocity field perturbations are phase-shifted by roughly a quarter wavelength, suggesting a breathing mode that is propagating in the direction of Galactic longitude $l\sim 270$~deg. The second feature is a gradient in the stellar number density and mean vertical velocity with respect to Galactocentric radius. This feature, which extends across the entire region of our analysis, may be associated with the extension of the Galactic warp into the Solar neighbourhood in combination with more localised bending waves.
}

   \keywords{Galaxy: kinematics and dynamics -- Galaxy: disk -- solar neighborhood -- Astrometry}

   \maketitle
%

\section{Introduction}

\emph{Gaia} Data Release 3 ({\emph Gaia} DR3) provides the measurements necessary to model the six-dimensional phase space distribution function (DF) of stars within a few kiloparsecs of the Sun \citep{GaiaMission2016, GaiaDR3, GaiaAsymmetry2022} and thereby test our understanding of stellar dynamics and galactic evolution. However, analysis of the full phase space DF is challenging due to the curse of dimensionality and the difficulty in visualising structure in six dimensions. In this paper, we focus on the stellar number density $n$ and mean vertical velocity $\overline{W}$, which are derived from velocity moments of the DF. Our interest in these quantities stems from recent observations of dynamical features in the disk associated with structure perpendicular to the Galactic mid-plane. These include the number count asymmetry about the mid-plane \citep{widrow2012,yanny2013,bennett2019,everall-II}, bending and breathing motions \citet{widrow2012,williams2013,carlin2013,GaiaDR2Kinematics2018,GaiaAsymmetry2022}, and the phase-space spiral in the $(Z,W)$-plane \citep{antoja2018}. There is also evidence that the disk is corrugated beyond the Solar circle from observations of both $n$ \citep{xu2015} and $\overline{W}$ \citep{schonrich2018,friske2019}.

In his seminal work on the vertical structure of the Milky Way, \citet{oort1932} combined observations of $n$ and $\overline{W}$ in the Solar neighborhood to infer the vertical acceleration $a_z$ as a function of $Z$. The essence of the calculation can be understood from dimensional considerations; near the mid-plane $a_z \simeq Z \left (\Delta W / \Delta Z\right )^2\simeq Z\Omega_z^2$, where $\Delta W$ and $\Delta Z$ are characteristic widths of the disk in $Z$ and $W$ and $\Omega_z$ is the frequency of vertical oscillations. In principle, simultaneous measurements of the spatial and velocity distributions associated with a disturbance of the disk would similarly allow one to study its dynamics. For example, armed with measurements of the displacement of the mid-plane and mean vertical velocity, one might be able to test theories of bending modes in the disk (see \citealt{sellwood2013} and references therein).

To date, most of the dynamical features mentioned above have been detected in either $n$ or $W$. One notable exception is the phase spiral, which is seen in number counts in the $(Z,W)$ phase space. The phase spiral almost certainly arises from the incomplete phase mixing of a perturbation to the disk and the presence of information in both $Z$ and $W$ allows one to date the perturbation (e.g. \citealt{antoja2018, laporte2019, li2021, widmark2022}). One of the main goals of this paper is to find other examples of perturbations that can be identified in both $n$ and $W$ and combine these two spatially varying fields in order to learn about the time-varying nature of disk perturbations.

The main challenge in inferring $n$ comes from understanding the selection effects. In particular, the \emph{Gaia} selection function is highly complex \citep{Boubert-II, everall-I, Everall-V,rybizki2021, CantatGaudin2022}; it depends mainly on stellar crowding and \emph{Gaia}'s scanning law and brightness limits, with the additional confounding issues of dust reddening and extinction. These different factors have a spatial dependence, but are independent with respect to velocity. Modelling significant selection effects is often challenging, making it difficult to accurately extract the stellar number density distribution. By comparison, it is more straightforward to derive the mean velocity field as a function of spatial position (e.g. \citealt{GaiaDR2Kinematics2018,2022MNRAS.512.1574M}) since the spatially dependent selection does not induce a strong systematic bias but only lowers the amount of available data. On the other hand, velocity measurements require parallax, proper motion, and radial velocity measurements. At present, the number of stars in \emph{Gaia} where this is possible is a factor of $\sim 30$ less than the total number of stars in the survey.

We modelled the spatial distribution of stars in the Galactic disk within a distance of roughly two kiloparsecs using data from \emph{Gaia} DR3 \citet{GaiaDR3}, supplemented with photo-astrometric distances from StarHorse \citep{2022A&A...658A..91A}. We assumed that the three-dimensional stellar number density distribution is a Gaussian process (GP) and used GP regression to estimate the underlying smooth number density field in a non-parametric way that does not rely on any symmetry assumptions such as axisymmetry or mirror symmetry about the Galactic plane. We carefully masked any spatial region that is compromised by a large distance, dust extinction, or the presence of open clusters. Thanks to the inherent property of smoothness of GPs, a masked spatial volume is still informed by its unmasked spatial neighbourhood. In this manner, we were able to construct a model-independent yet robust three-dimensional map of the stellar number density distribution within a distance of a few kiloparsecs. We also mapped the mean vertical velocity field $\overline{W}$ using a similar approach, for example masking open clusters, although we simply calculated the mean value in fixed spatial volumes without any GP regression. For a GP model of the velocity field, see \citet{nelson2022}.

This article is structured as follows. In Sect.~\ref{sec:data}, we present the data and define our coordinate system. We describe our method for mapping the stellar number density distribution in Sect.~\ref{sec:nu_dist}, and our method for mapping the vertical velocity distribution in Sect.~\ref{sec:vel_dist}. In Sect.~\ref{sec:results}, we present our results. In the final Sects.~\ref{sec:discussion} and \ref{sec:conclusion}, we discuss and conclude.

\section{Data}\label{sec:data}

\begin{figure}
	\centering
	\includegraphics[width=1\columnwidth]{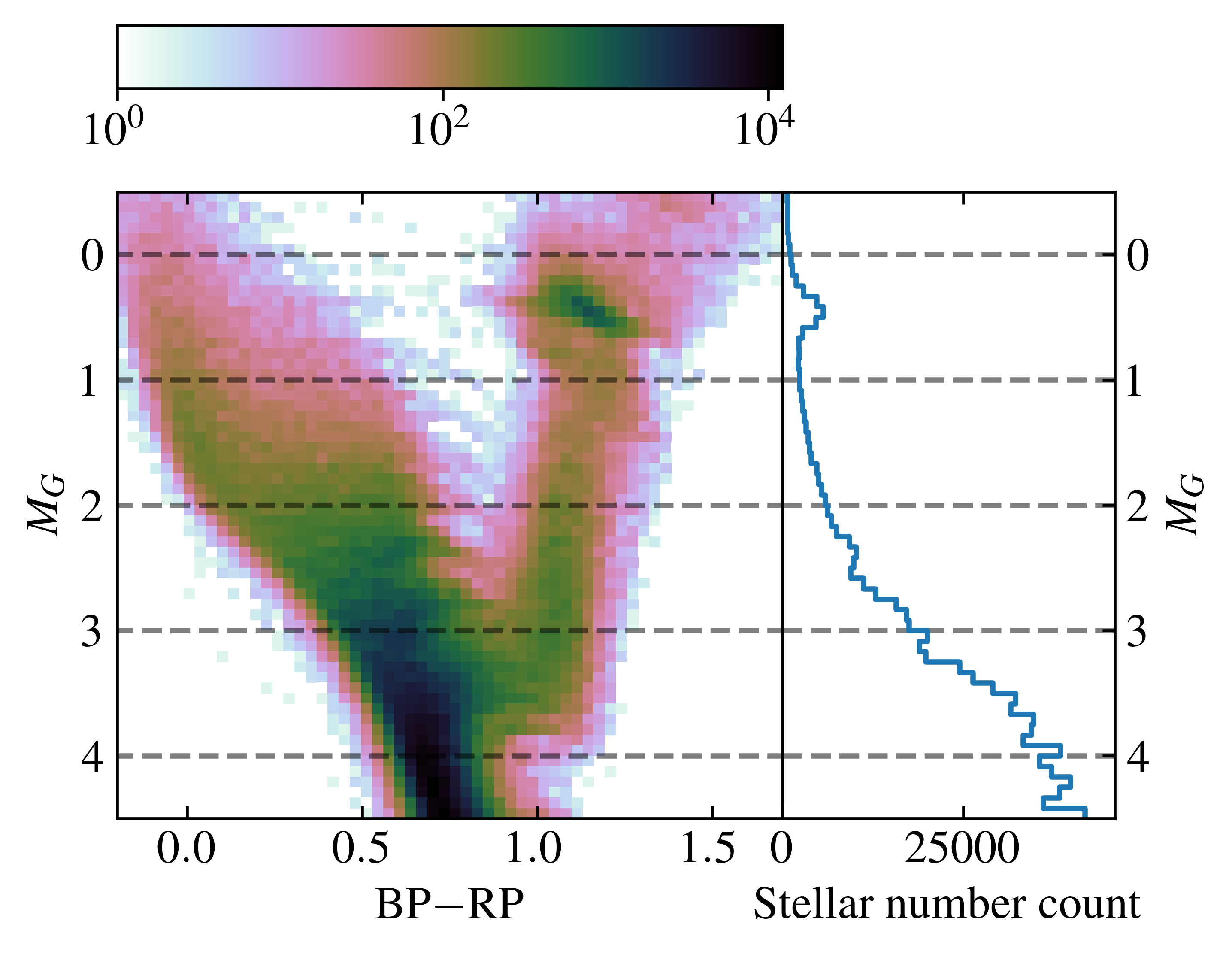}
	\caption{Colour magnitude diagram of stars in StarHorse within a distance of 400~pc. The panel on the right side shows the one dimensional absolute magnitude histogram. The dashed lines correspond to the magnitude cuts of our four data samples. The colour and absolute magnitude values are intrinsic and dust-corrected as given directly by the StarHorse catalogue.}
	\label{fig:CMD}
\end{figure}

We used data from \emph{Gaia} DR3, supplemented with photo-astrometric distance and dust extinction information from StarHorse \citep{2022A&A...658A..91A}, which is available for \emph{Gaia} stars with an apparent magnitude $m_G < 18.5$. We analysed four different stellar populations defined by a range in absolute magnitude in the \emph{Gaia} $G$-band according to $M_G \in (0, 1],\, (1, 2],\, (2, 3],\, (3, 4]$. A colour magnitude diagram illustrating our data sample cuts can be found in Fig.~\ref{fig:CMD}. In this paper, all colours and absolute magnitudes (but not apparent magnitudes) are taken directly from the StarHorse catalogue, and thus include a correction for dust reddening or extinction. We plot the age distribution of the respective stellar samples in Appendix~\ref{app:ages}.

We used the Cartesian heliocentric coordinates $\boldsymbol{X} = (X,Y,Z)$, which point in the directions of the Galactic centre, Galactic rotation, and Galactic north. In terms of the Galactic longitude and latitude, written $l$ and $b$, these coordinates are defined according to
\begin{equation}
    \begin{split}
        X & = d \, \cos l \, \cos b, \\
        Y & = d \, \sin l \, \cos b, \\
        Z & = d \, \sin b, \\
    \end{split}
\end{equation}
where $d$ is the distance from the solar position. We also made use of the Galactocentric cylindrical radius, given by
\begin{equation}
    R = \sqrt{(R_\odot-X)^2+Y^2},
\end{equation}
where we assumed a value of $R_\odot = 8.2~\kpc$ for the Sun's distance from the Galactic centre (consistent, for example, with \citealt{mcmillan2016mass}).

The time-derivatives of these spatial positions, $d\boldsymbol{X}/ \de t$, give the velocities in the solar rest-frame. In this work, we focused on the vertical velocity, which is given by
\begin{equation}
    W = \de Z/ \de t = d \, k_\mu \, \mu_b \, \cos b + v_\text{RV} \, \sin b
\end{equation}
where $k_\mu = 4.74057 ~ \yr \, \mas^{-1} \, \kpc^{-1} \, \kmsec$ is a unit conversion constant, $\mu_b$ is the latitudinal proper motion, and $v_\text{RV}$ is the radial velocity.

We accounted for the statistical uncertainties in number counts while neglecting observational uncertainties in the positions of individual stars. For a star's spatial positions, we used the \emph{Gaia} DR3 values for Galactic latitude and longitude and the StarHorse median value for the distance (labelled \texttt{dist50} in that catalogue). When calculating the velocities, we used a similar procedure, neglecting observational uncertainties for the proper motions and radial velocity, although with the data quality cuts described in Sect.~\ref{sec:vel_dist}.

The relative precision of StarHorse distances is 3~\% at the bright end of the luminosity function but only 15~\% for $m_G\sim 17$  (see figure 13 in \citealt{2022A&A...658A..91A}). The spatial volume we studied is potentially problematic due to the high rate of dust extinction and stellar crowding. At a distance of a few kiloparsecs, even a relative uncertainty of a few per cent is significant for our purposes, especially where there are strong degeneracies between distance and dust extinction. The data cuts we applied in order to circumvent these issues are described below in Sect.~\ref{sec:masks}.

\section{Stellar number density distribution}\label{sec:nu_dist}

For each of our four stellar populations, defined by different cuts in absolute magnitude, we carefully masked spatial volumes that were compromised by large distances, high dust extinction, or the presence of open clusters. In the remaining spatial volume, where we could consider the data sample to be complete, we fitted a three-dimensional stellar number density distribution function using a GP. These steps are described in detail below.

\subsection{Masks}\label{sec:masks}

Our strategy was to choose a spatial volume and range in apparent magnitude that minimised completeness issues. Specifically, we limited ourselves to $m_G$ in the range of 6--18, where the \emph{Gaia} completeness function is close to unity, typically with deviations that are a few per cent at most \citep{everall2022,CantatGaudin2022}. Furthermore, we masked areas of the sky where the number density is biased by the presence of open clusters. For this purpose, we constructed a mask function, written $\text{mask}(\boldsymbol{X})$, which can either take a value of either zero or unity at every point in three-dimensional space. Any unmasked spatial volume was assumed to be complete.

We began by constructing a three-dimensional dust extinction map as a function of the angles $l$ and $b$
and the cylindrical radius $R_\text{cyl} = \sqrt{X^2 + Y^2}$. We divided the $(l,b)$ sky using a HEALPix map of order 7 (corresponding to an angular resolution of 0.46~degrees)
and divided $R_\text{cyl}$ into segments of width 100 pc. Each combination of $R_\text{cyl}$ segment and $(l,b)$-pixel corresponded to its own spatial volume for which we calculated the 80th percentile of dust extinction using StarHorse data (column \texttt{ag50}) for all stars with $M_G<6~\magn$. Although we binned in terms of $R_\text{cyl}$, the map can equally well be understood in terms of heliocentric distance with a resolution in distance that depends on $b$.

Within each of these three-dimensional $(l, b, R_\text{cyl})$ volumes, the mask function was set to unity only where it was fulfilled that
\begin{equation}\label{eq:appGbound}
M_{G,\text{high}}+5 \, \text{log}_{10}\left( \frac{\text{distance}}{10~\pc} \right)
+ (\text{80th percentile dust ext.})<17.
\end{equation}
This criterion ensured that the non-masked data of some given spatial position has a distribution of apparent magnitudes which falls largely below 17, with only a weaker tail of stars that were dimmer than this limit. In this sense, using the 80th dust extinction percentile is a conservative measure.

We also masked the spatial volume where open clusters affected the stellar number density or obscured the field of view. We used the catalogue of open clusters from \cite{2018A&A...618A..93C}. For each open cluster, we assumed an angular size given by two times its half-light radius (\texttt{r50} in the open cluster catalogue). Using the same sky map as defined for the dust mask above, we masked any spatial volume where the $(l,b)$-pixel overlapped with an open cluster's angular area and the spatial distance extended beyond the 5th percentile distance of the open cluster (\texttt{d05} in the open cluster catalogue). In doing so, we masked the spatial volumes that lie behind the sight-line of an open cluster, thereby mitigating incompleteness effects that may arise from stellar crowding.

\begin{figure}
	\centering
	\begin{subfigure}{1\columnwidth}
	    \includegraphics[width=0.94\columnwidth]{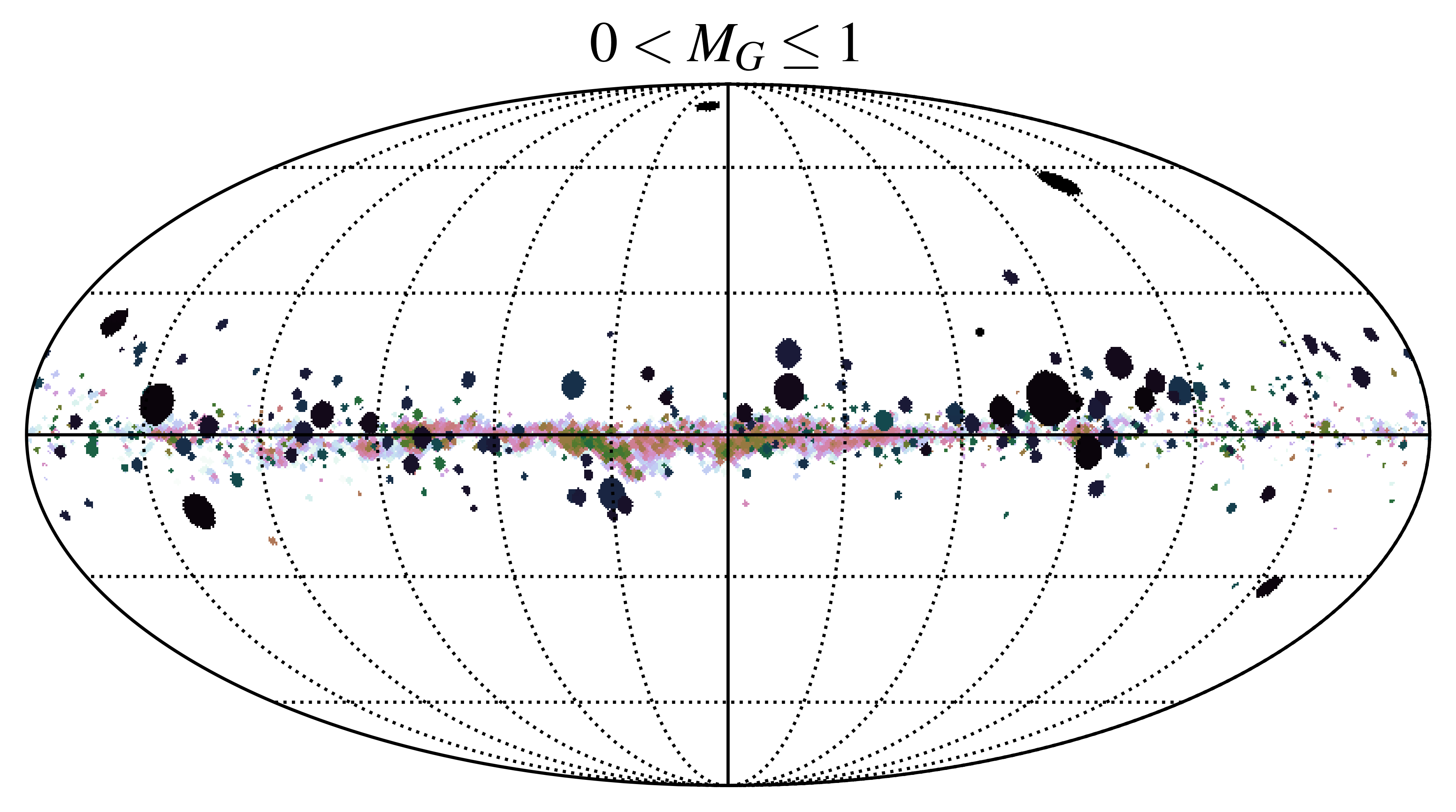}
	\end{subfigure}
    \newline
    \begin{subfigure}{1\columnwidth}
	    \includegraphics[width=0.94\columnwidth]{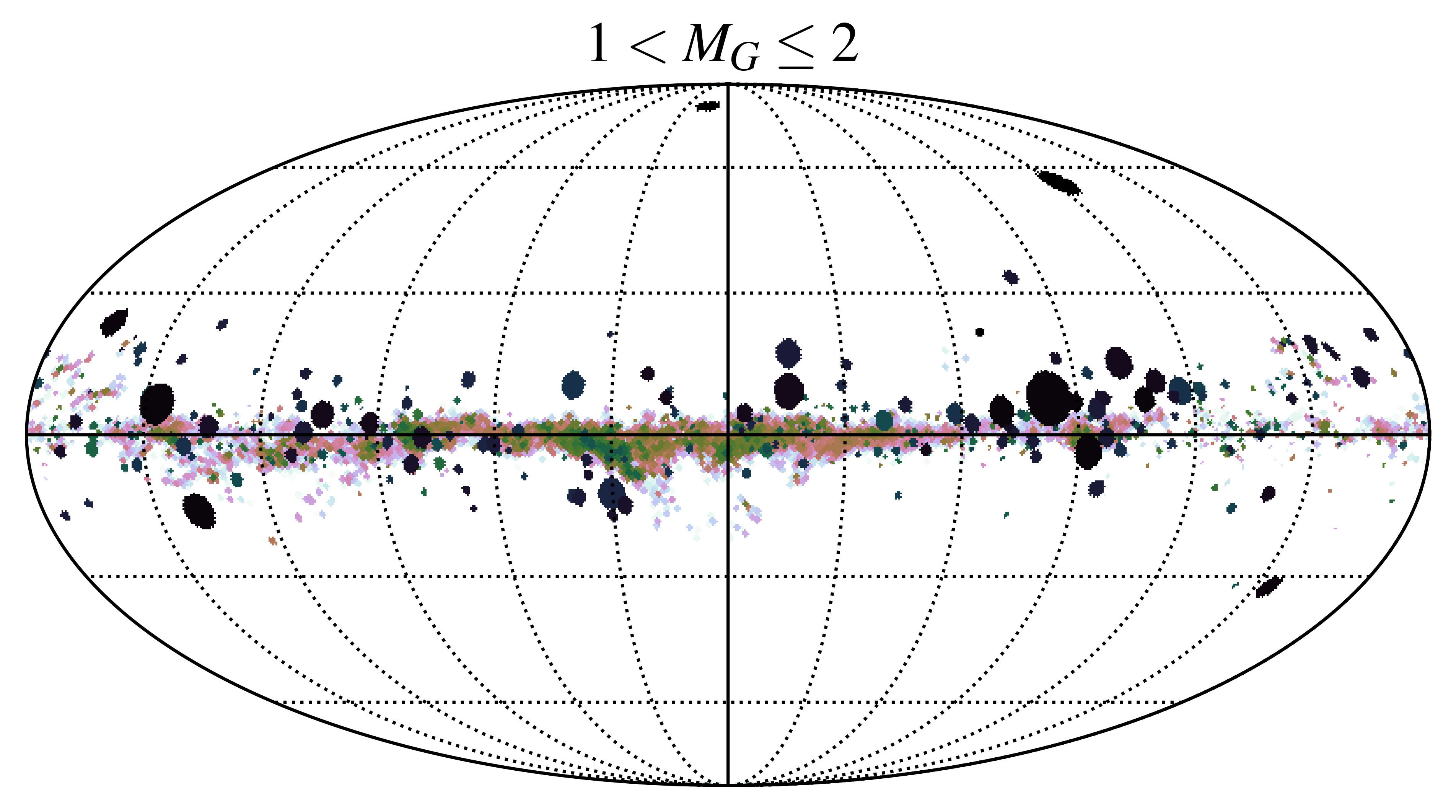}
	\end{subfigure}
    \newline
    \begin{subfigure}{1\columnwidth}
	    \includegraphics[width=0.94\columnwidth]{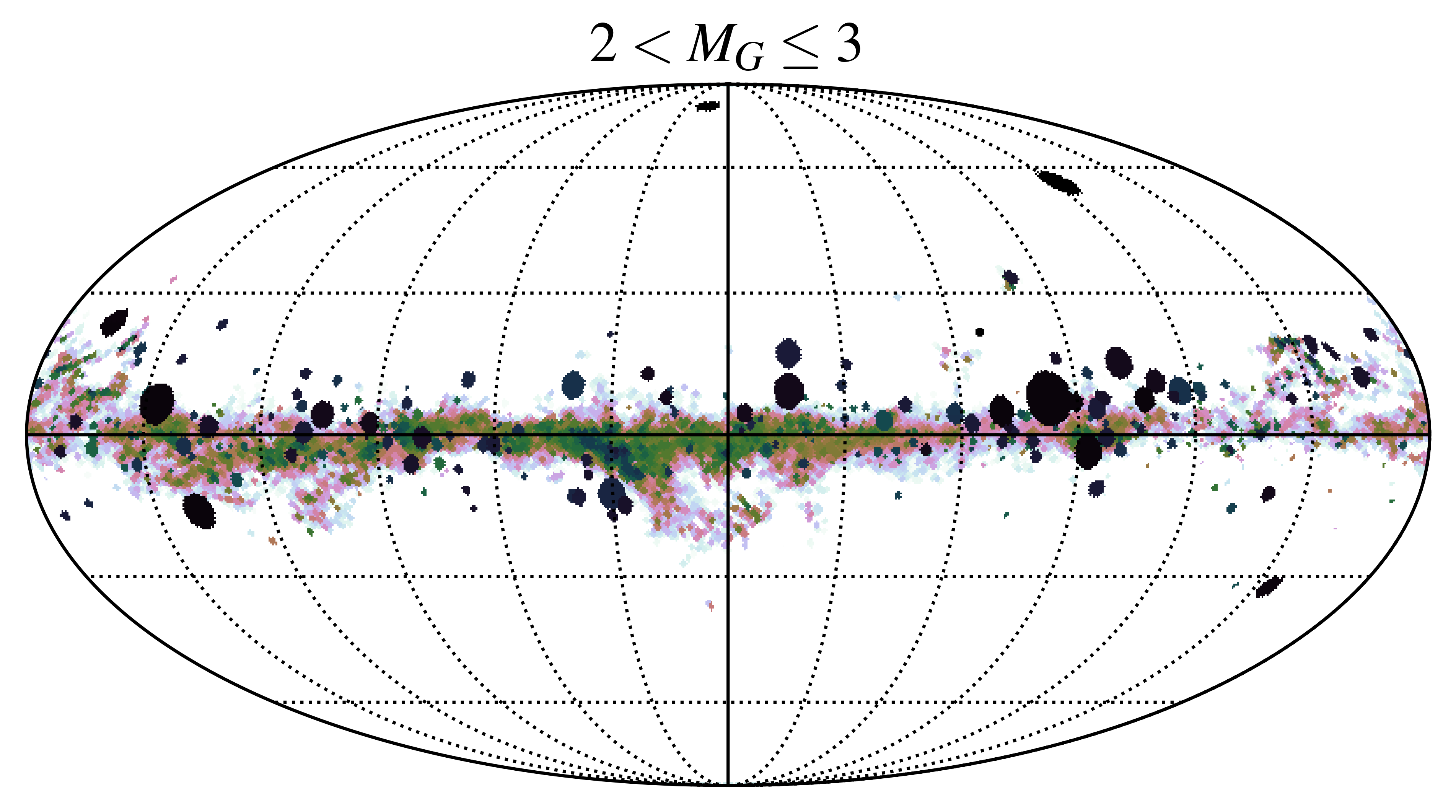}
	\end{subfigure}
    \newline
    \begin{subfigure}{1\columnwidth}
	    \includegraphics[width=0.94\columnwidth]{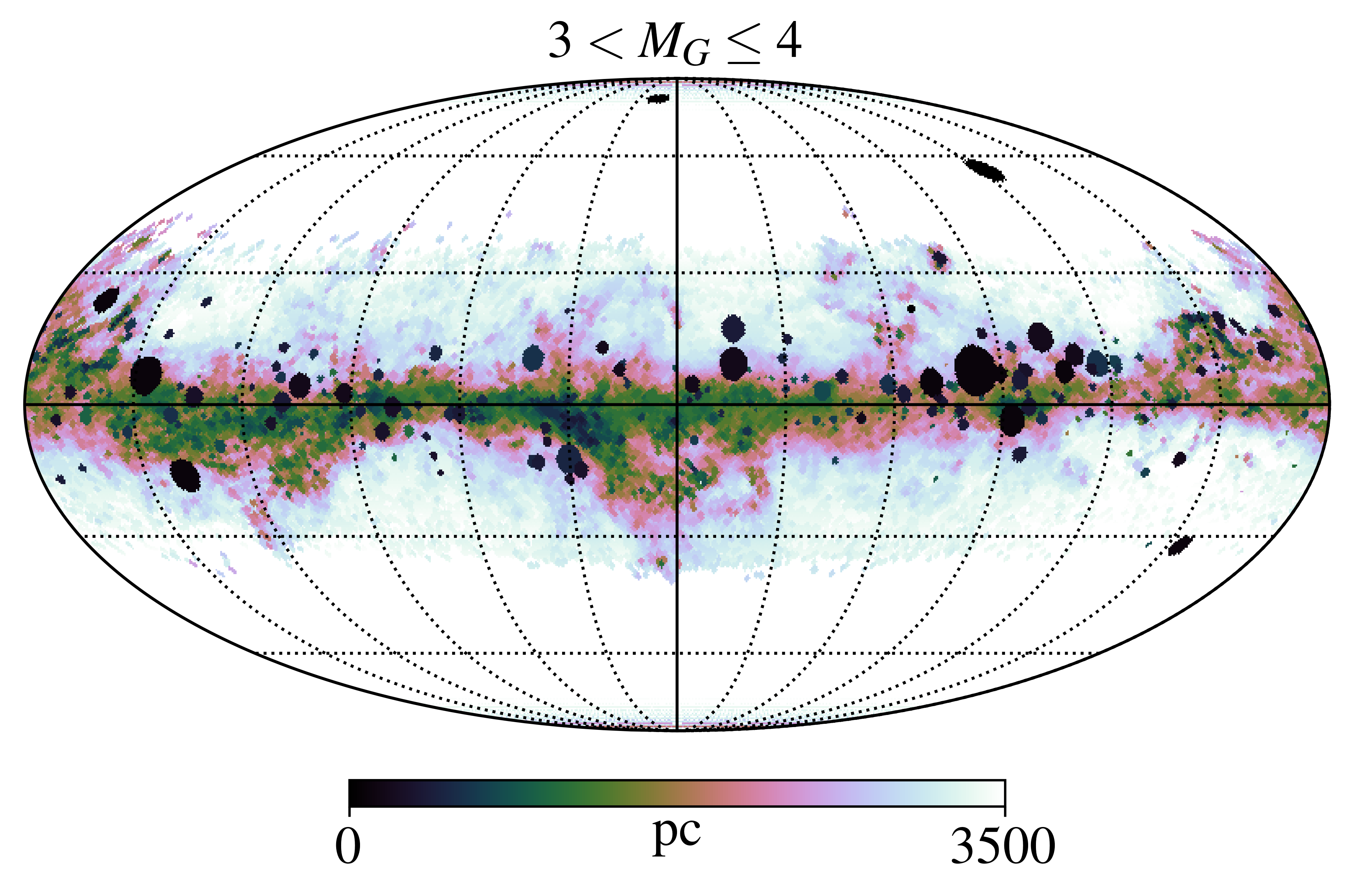}
	\end{subfigure}
	\caption{Upper distance limit, given by Eq.~\ref{eq:appGbound}, as a function of angles $l$ and $b$, for our four data samples. The centre of the map is in the direction of the Galactic centre, while positive $b$ is pointing upwards and positive $l$ is pointing to the right.}
	\label{fig:masks}
\end{figure}

The mask functions of our stellar samples can be seen in Fig.~\ref{fig:masks}. The circular patches are masked due to open clusters, while the remaining more complex structure arises from dust extinction and the limit in apparent magnitude as defined in Eq.~\ref{eq:appGbound}. The four data samples shown in the figure differ in their spatial extent, where the brightest one reaches greater distances.

Apart from these upper distance constraints, we also masked the nearby spatial volume in order to avoid stars that are too bright. We set a lower limit in distance, requiring that this criterion was fulfilled:
\begin{equation}
    M_{G,\text{low}} + 5 \log_{10} \left(\frac{\text{distance}}{10~\pc}\right) > 6,
\end{equation}
where $M_{G,\text{low}}$ is the lower absolute magnitude bound of the data sample. This distance limit, in combination with the cuts in absolute magnitude, ensures that a star in our sample could never be brighter than $m_G = 6$, regardless of dust extinction.

\subsection{Number count and number density}\label{sec:number_density}

The data for our GP analysis were reduced in the following way. We divided the spatial volume into a three-dimensional Cartesian grid with a 100~pc spacing in $X$ and $Y$, and a 10~pc spacing in $Z$. We chose a smaller grid spacing in the $Z$ direction, since the scale length for variations normal to the Galactic plane is smaller than the scale length for variations parallel to the Galactic plane. Furthermore, because we wanted to study the Galactic disk, we restricted ourselves to $|Z| \leq 800~\pc$. Each volume cell, written $V_{i,j,k}$, was labelled by the triplet of indices $(i,j,k)$, which sets its spatial boundaries according to
\begin{equation}\label{eq:XYZcells}
    \begin{split}
        100\times i -50 <  & \frac{X}{\pc} \leq 100\times i +50, \\
        100\times j -50 < & \frac{Y}{\pc} \leq 100\times j +50, \\
        10\times k < & \frac{Z}{\pc} \leq 10\times (k+1).
    \end{split}
\end{equation}
These volume cells could be partly or completely masked by the mask function described above.

The stellar number count in a given volume cell was related to its number density in the following way. Each volume cell had its specific number count, written $N_{i,j,k}$, given by the number of stars that remained in the volume cell after applying the mask function. We also associated each volume cell with an effective volume, written $\Omega_{i,j,k}$, corresponding to the non-masked volume within that cell:
\begin{equation}
    \Omega_{i,j,k} =
    \int_{V_{i,j,k}} \text{mask}(\boldsymbol{X})\, \de^3 \boldsymbol{X}.
\end{equation}
The effective volume of each cell was calculated numerically via Monte Carlo integration, and took values in the range $[0,10^5]~\pc^3$. These quantities were of course unique for each separate data sample.

For each volume cell that was not completely masked, the stellar number density was given by
\begin{equation}\label{eq:normed_N}
    n_{i,j,k} = \frac{N_{i,j,k}}{\Omega_{i,j,k}}.
\end{equation}
We took the associated statistical uncertainty of $n_{i,j,k}$ to be
\begin{equation}\label{eq:sigma_ijk}
    \sigma_{i,j,k} =
    \frac{(N_{i,j,k}^2+5^2)^{1/4}}{\Omega_{i,j,k}},
\end{equation}
which corresponds to a Poisson count uncertainty in the limit of high data counts. We added a number 5 in quadrature in the nominator in order to decrease the statistical power where the data count is very low. Because we estimated the uncertainty from the data count, rather than from an underlying model that generates it, this statistical uncertainty was often underestimated especially for data bins with low number counts. We were mainly interested in the results where the number count was fairly high and the added number 5 was negligible; however, adding this number was necessary in order to avoid fitting artefacts, due to very low number count values at large distances from the Galactic mid-plane or where $\Omega_{i,j,k}$ was close to zero. Choosing a slightly different number did not significantly alter our results.

\begin{figure*}
	\centering
	\includegraphics[width=1.\textwidth]{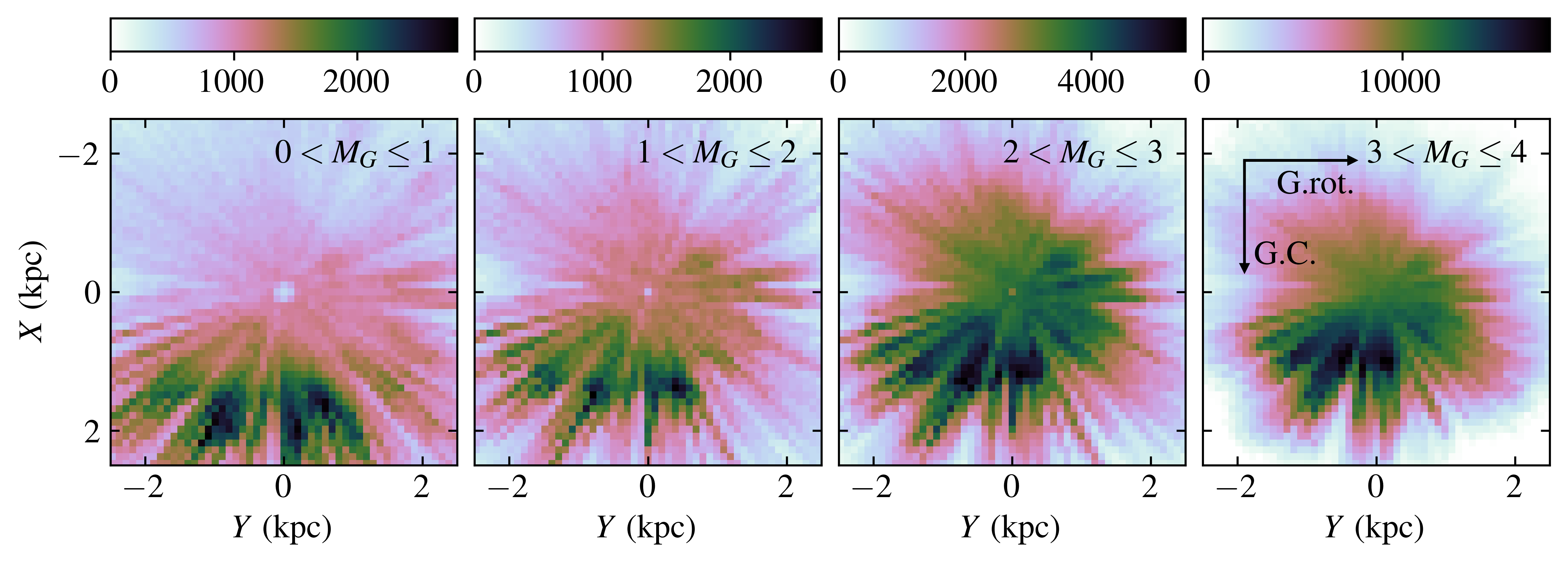}
	\caption{Stellar number counts per area cell in the $(X,Y)$-plane, for our four data samples (specified in the panels' top right corners), after masks have been applied. The arrows in the rightmost panel show the direction of the Galactic centre and the direction of Galactic rotation. The axis ranges are shared between all panels.}
	\label{fig:number_counts}
\end{figure*}

The disk plane projections of number counts, after masks had been applied, can be seen in Fig.~\ref{fig:number_counts}. The total number of volume cells that were not completely masked (i.e. $\Omega_{i,j,k}>0~\pc^3$) was equal to 1\,005\,181, 806\,270, 584\,684, 310\,122 for our four data samples (going from brightest to dimmest). The total number of stars in the non-masked spatial volume are 4\,613\,344, 3\,307\,223, 6\,820\,434, 13\,854\,829.

\subsection{Gaussian process fit}\label{sec:GP}

In this section, we describe how we modelled the normalised number counts as a Gaussian Process (GP). GP methods allow one to infer or interpolate an underlying function given a finite number of function observations. The main attraction of GPs for this work is that they allowed us to model the stellar number density $n(\boldsymbol{X})$ as a smooth and differentiable function without imposing a parametric form that presupposes constraints such as Galactic axisymmetry. In addition, data uncertainties can be incorporated into GP modelling as long as they are approximately Gaussian. 

Formally, a GP is a collection of random variables with the property that any finite subset of these variables has a multivariate normal distribution (see for example \citealt{10.5555/1162254}). The probability distribution function (PDF) for ${\cal N}$ random variables from a GP is therefore defined by an ${\cal N}\times {\cal N}$ covariance matrix. In our case, the variables were the normalised number counts as labelled by $i,j,k$, while the elements of the covariance matrix depended on the distances between pairs of volume bins through a function usually called the kernel. In this paper, we used the radial basis function kernel, guaranteeing continuity and smoothness, for which the covariance matrix element for two bins with grid indices $(i,j,k)$ and $(i',j',k')$ is
\begin{equation}\label{eq:kernel}
    k({\bf X}_{i,j,k},\,{\bf X}_{i',j',k'}) = A
     e^{-(X_i-X_{i'})^2/l_x^2}
    e^{-(Y_j-Y_{j'})^2/l_y^2}
    e^{-(Z_k-Z_{k'})^2/l_z^2}~.
\end{equation}
The parameters $A$, $l_x,\,l_y,\,$ and $l_z$ determine the overall variance and length scales associated with structure in the number counts. Proper choice of these hyperparameters is essential to finding a suitable model. 

Suppose we want to infer the number counts at some new position ${\bf X}^*$. The joint PDF for ${\cal N}$ data points and the new point is defined by an $({\cal N}+1)$-dimensional Gaussian. On the other hand, the conditional PDF for the new point given the data is found by marginalising over the data via Bayes theorem. Since the PDFs are all Gaussian, the marginalisation integrals can be done analytically. However, for this step, one must invert the $N\times N$ data covariance matrix, which is an ${\cal O}(N^3)$ operation that requires ${\cal O}(N^2)$ of rapid-access memory.

An exact GP analysis of our full data set was unfeasible given the large number of measurements and the CPU and RAM requirements of the GP calculation. There are numerous approximation schemes such as the inducing point method that allow one to apply GP regression to very large data sets (see \citealt{Titsias2009} and references therein). Here, we took the simple approach of applying GP regression to smaller spatial sub-volumes, rather than to the complete spatial volume all at once. For each area cell in the $(X,Y)$-plane, we fitted a new GP to its surroundings, including all other area cells within 650~pc (i.e. $i_\text{diff.}^2+j_\text{diff.}^2<6.5^2$). The GP was fitted to the normalised number count $\tilde{N}$, with its associated uncertainty $\tilde{\sigma}$.

In terms of the hyperparameters of the GP, as expressed in Eq.~\eqref{eq:kernel}, we set the variance $A$ to be equal to the variance of the normalised number count in the non-masked volume cells and used the spatial correlation scale lengths $(l_x,l_y,l_z) = (300, 300, 100)~\pc$. Due to the computational cost and the shortcut of implementing GPs in sub-volumes, we did not attempt to fit the hyperparameters. Even if fitting the hyperparameters would be computationally feasible, it still might not be desirable. The hyperparameters were specifically chosen such that the stellar number density fits would have certain properties of being correlated over reasonable spatial scales. Choosing a smaller correlation scale lengths could make it too sensitive to perturbations and systematic issues on smaller spatial scales. For example, there are degeneracies between parallax and absolute magnitude, as well as with the three-dimensional distribution of dust, giving rise to spatially correlated systematic errors, potentially on smaller scales. Our choice is further discussed and motivated in the beginning of Sect.~\ref{sec:results}.

\subsection{Symmetric analytic function}\label{sec:symm_fit}

For our non-parametric GP fit to the data, we were mainly interested in the perturbations with respect to some smooth background, and in what ways the symmetries of the Galactic disk are broken. In order to study such residuals, we also performed a parametric fit to our data using an analytic stellar number density distribution function, which was fully axisymmetric and mirror symmetric across the mid-plane. For this purpose, we used a mixture model of three disk components which take a functional form
\begin{equation}\label{eq:f_symm}
\begin{split}
    & f_\text{symm.}(R,Z \, | \, a_i,L_i,H_i,Z_\odot) = \\ 
    & \sum_{i=1}^{3}
    a_i \, \exp \left(-\frac{R-R_\odot}{L_i} \right) \, \text{sech}^2
    \left( \frac{Z+Z_\odot}{H_i} \right).
\end{split}
\end{equation}
It has ten free parameters: $a_i$ are the respective amplitude of the three disk components; $L_i$ are their scale lengths; $H_i$ are their scale heights; and $Z_\odot$ is the height of the Sun with respect to the disk mid-place. We constrained $a_i$ to be positive, $L_i>500~\pc$, and $H_i>100~\pc$.

We fitted $f_\text{symm.}$ to the measured stellar densities in the non-masked spatial volume by maximising a Gaussian likelihood with the Adam optimiser \citep{adamopt}. We used the same normalised stellar number count and statistical uncertainty as are defined in Eqs.~\eqref{eq:normed_N} and \eqref{eq:sigma_ijk}, to the spatial volume that includes all area cells where the mean effective fractional volume for $|Z|<500~\pc$ was larger than 50~\%. For each of our four data samples, we performed separate fits of $f_\text{symm.}$. The fitted parameters are found in Appendix~\ref{app:symm_fit_res}, where we also discuss some additional tests (e.g. fitting a smaller or larger number of disk components).

The main purpose of this function is to facilitate the visualisation of the GP model and serve as a smooth and symmetric background distribution for comparison purposes. For this reason, we refrain from making any strong physical interpretation of this function in isolation.

\section{Vertical velocity distribution}\label{sec:vel_dist}

In addition to the stellar number density field, we also studied the vertical component of the velocity field. We calculated the mean vertical velocity of our four stellar samples, as a function of spatial position. We did so from the radial velocity sample, requiring a radial velocity uncertainty smaller than $5~\kmsec$. The vertical velocity of each such star was given directly by its StarHorse distance (\texttt{dist50}) and \emph{Gaia} DR3 velocity information (neglecting observational uncertainties). We also produced results with stronger data quality cuts in both proper motion and distance uncertainty, but saw only small differences in the results.

We cleaned the data of open clusters. For each open cluster, we masked the spatial volume defined by an angular radius within $3\times\texttt{r50}$ of its sky angular position and a distance from the Sun in range $(\texttt{d05}-3\times \texttt{r50}$,\, $\texttt{d95}+3\times \texttt{r50})$, where $\texttt{r50}$ is the half-light radius and $\texttt{d05}$ ($\texttt{d95}$) is the 5th (95th) distance percentile in the open cluster catalogue of \cite{2018A&A...618A..93C}. Hence, this open cluster mask was slightly different from the one applied when studying the stellar number density field, where we also masked any spatial volume that lies behind the open cluster. For the vertical velocity field, open clusters are problematic because they are not representative of the bulk stellar distribution, while incompleteness effects that arise due to stellar crowding behind an open cluster are not expected to produce a significant bias.

We divided the disk plane using the same area cells as defined in Eq.~\ref{eq:XYZcells}. We divided the bins in terms of height, using bin edges at 0, 50, 100, 200, 300, 500, and 700 pc for the Galactic north, and the corresponding negative values for the Galactic south. For the transformation to the disk rest frame, we used a fixed value of $Z_\odot=15~\pc$ for all data samples. The total number of stars with velocity information are 1\,912\,727,
786\,600,
1\,353\,870, and
2\,243\,873, for our four data samples.

\section{Results}\label{sec:results}

\begin{figure*}
	\centering
	\includegraphics[width=.9\textwidth]{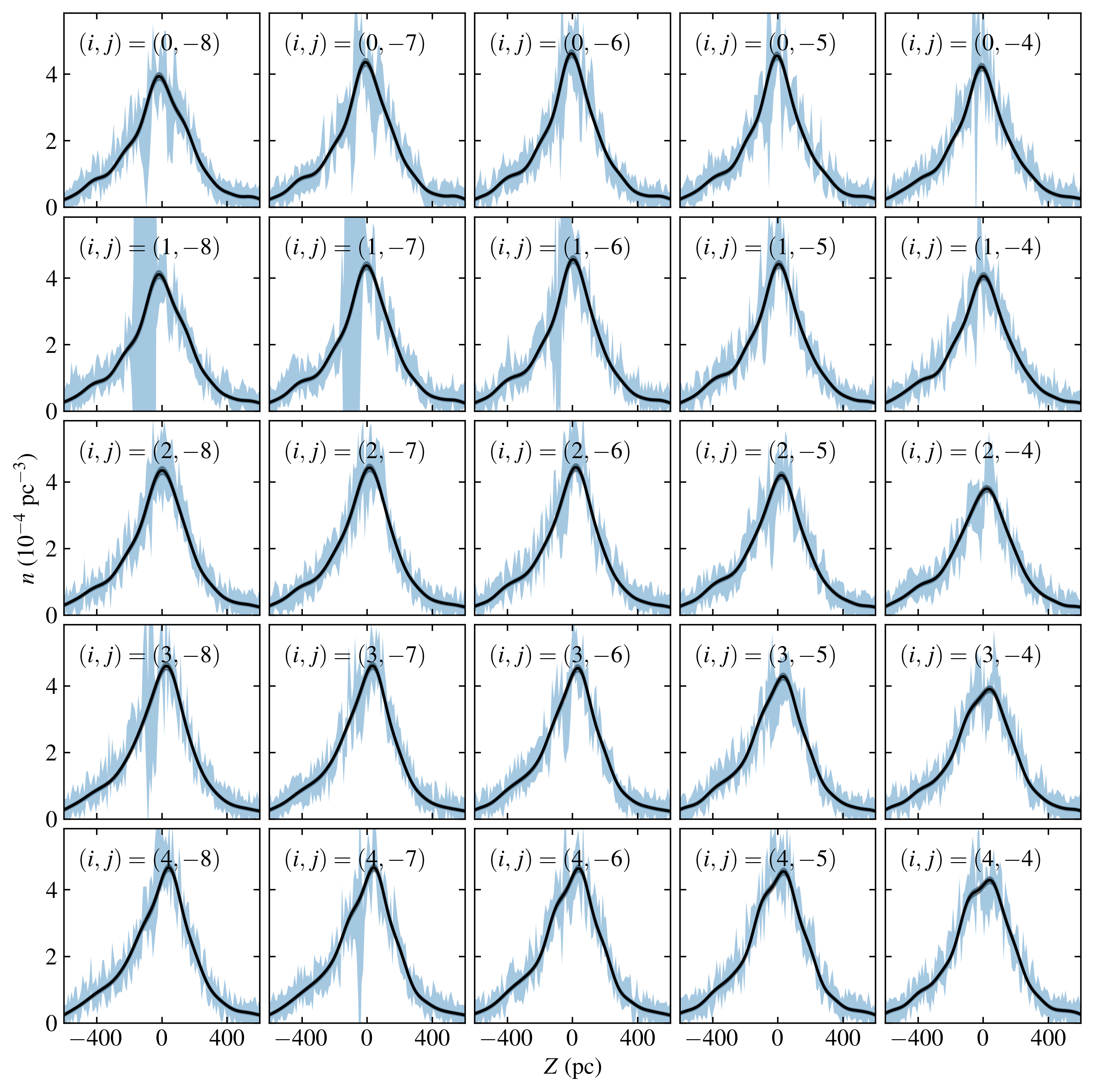}
    \caption{GP fit for data sample with absolute magnitude cuts $2 < M_G \leq 3$. Each panel corresponds to a 100-by-100 pc area cell in the $(X,Y)$-plane, labelled by indices $i$ and $j$ according to Eq.~\eqref{eq:XYZcells}, thus centred on $(X,Y)=(200,-600)~\pc$. The horizontal and vertical axes show height with respect to the Sun and the normalised stellar number count as defined in Eq.~\ref{eq:normed_N}. The solid lines correspond to the GP fits, with a smooth shaded region signifying its dispersion (mostly too small to see by eye). The jagged shaded region corresponds to the 1-$\sigma$ band of the data number count. The axis ranges are the same for all panels.}
    \label{fig:GPfit}
\end{figure*}

In Fig.~\ref{fig:GPfit}, we show the GP fit for a group of 25 neighbouring area cells.  This area of the $(X,Y)$-plane was chosen to illustrate a few key points. As can be seen in the second row first column panel, the presence of an open cluster has completely masked the number count information at $Z \simeq -100~\pc$. However, because the GP is correlated with nearby spatial regions, the fitted curve is still inferred in this sub-volume, with reasonable results. By comparing the fit in the respective panels, we can see that the fitted $n(Z)$ distribution varies somewhat in shape; for example, in some panels $n(Z)$ is clearly more skewed than in others. Our spatial correlation lengths of (300,300,100)~pc seem to be good choices. This conclusion was confirmed by trying other values, both greater and smaller. Our fit picks out interesting structures in $n(Z)$ on the hundred parsec scale. On the other hand, it smooths out smaller scale structures in the data, such as the feature near the mid-plane in the centre panel that has a spatial scale of a few tens of parsecs. We view these properties as an advantage of our method; structures that are considerably smaller than the disk scale height could well be artefacts of some systematic error, for example related to small scale structures in the dust distribution.
With that in mind, caution should be taken when interpreting these results, as they are a product of a specific data processing procedure and not a perfect or complete representation of the underlying data.

\begin{figure*}
	\centering
	\includegraphics[width=0.6\textwidth]{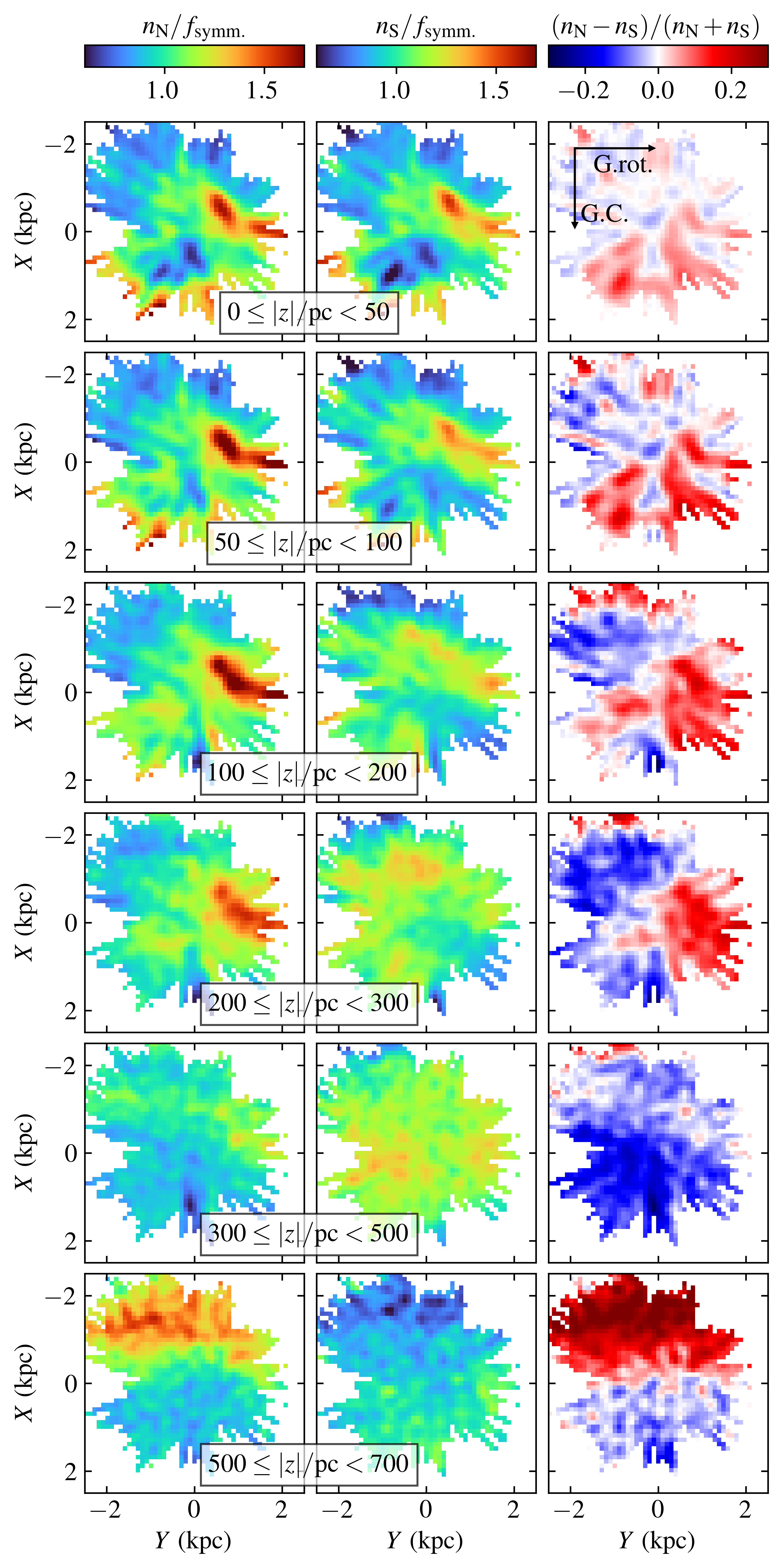}
    \caption{Stellar number density variations in the $(X,Y)$-plane of the data sample with $2<M_G\leq3$, for different bins in height. The left (middle) column shows the density variations north (south) of the mid-plane, as the ratio between the GP and symmetric analytic fit (as described in Sects.~\ref{sec:GP} and \ref{sec:symm_fit}, respectively). The right column shows the asymmetries between the north and south of the GP fits, where each row corresponds to a specific range in height with respect to the mid-plane's location when fitting $f_\text{symm.}$. The arrows in the top right panel show the directions of the Galactic centre and Galactic rotation. The axes ranges are shared between all panels.
    }
    \label{fig:disk_residuals_2<G<3}
\end{figure*}

In Figure~\ref{fig:disk_residuals_2<G<3}, we show the number density perturbations from the $2<M_G < 3$ data sample as projected onto the disk plane for different bins in $Z$. These perturbations are shown in terms of the ratio between our GP fit and the fitted symmetric function ($f_\text{symm.}$, as described in Sect.~\ref{sec:symm_fit}; its fitted parameters are found in Appendix~\ref{app:symm_fit_res}). There are a number of prominent perturbation features. First, there is an over-density at around $(X,Y)=(-0.3,0.8)~\kpc$ and for bins close to the mid-plane ($Z < 300~\pc$). The structure is fairly symmetric across the north and south and matches the location of the Local Spiral Arm found by \cite{Xu2013} and \cite{Reid2014,Reid2019}. Secondly, at greater heights, mainly for $500 \leq |Z| < 700~\pc$, there are strong asymmetries between the north and south density fields, roughly corresponding to a dipole oriented along the $X$-axis. Thirdly, there are asymmetries between the north and south mainly around the disk location $(X,Y)=(1,-1)~\kpc$ and $|Z| < 100~\pc$. This region is highly affected by dust extinction and stellar crowding and we cannot rule out the possibility that the feature is, at least in part, a systematic artefact.

\begin{figure}
	\centering
	\includegraphics[width=1\columnwidth]{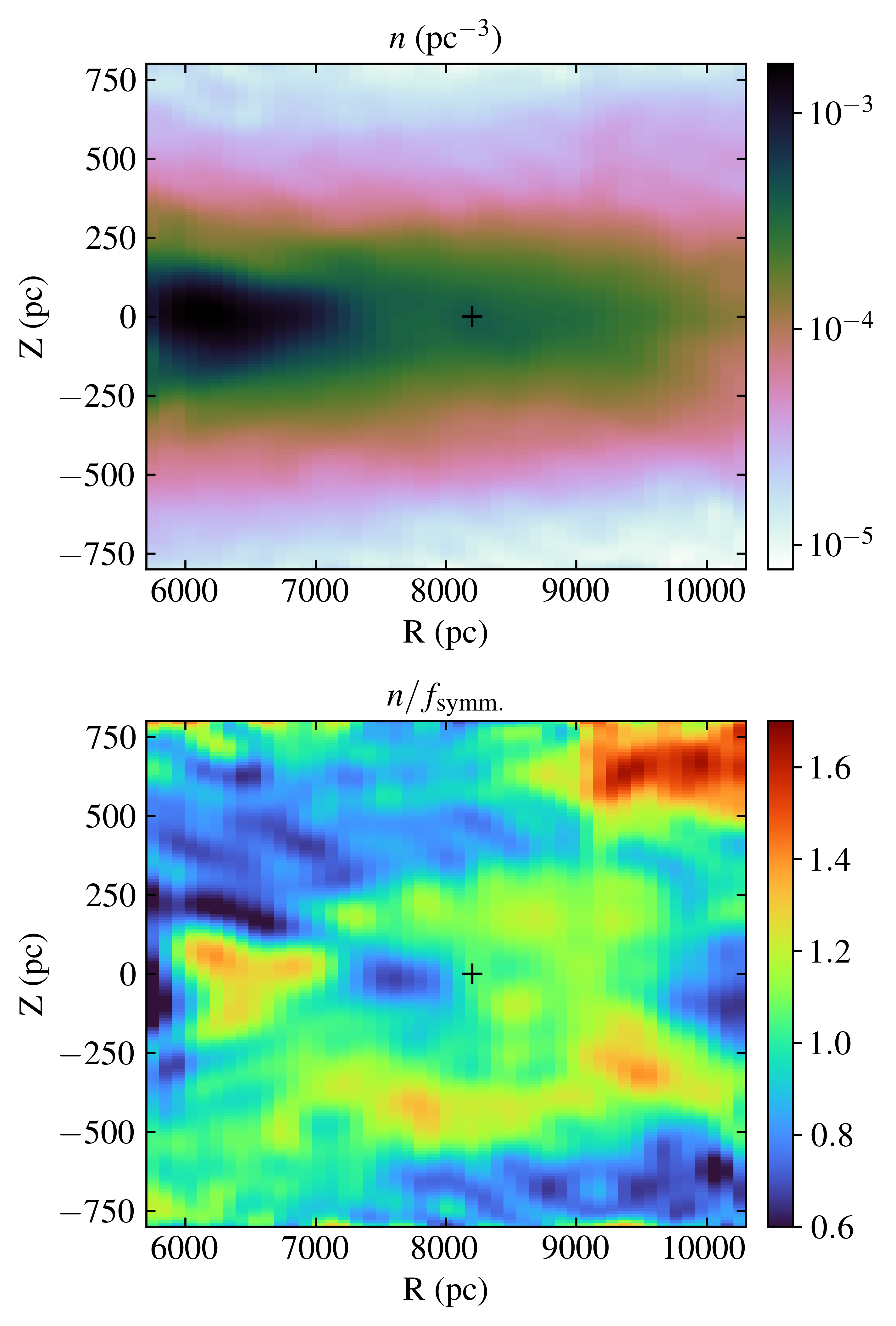}
    \caption{Stellar number count in the plane of Galactocentric radius and height, for data sample $2 < M_G \leq 3$, averaged over the spatial volume within $|Y|<250~\pc$. The top panel shows the number count of the GP fit, while the bottom panel shows the ratio with respect to the symmetric analytic fit. The Solar position is highlighted with a black plus marker.}
    \label{fig:Rz_residuals_2<G<3}
\end{figure}

\begin{figure}
	\centering
	\includegraphics[width=1\columnwidth]{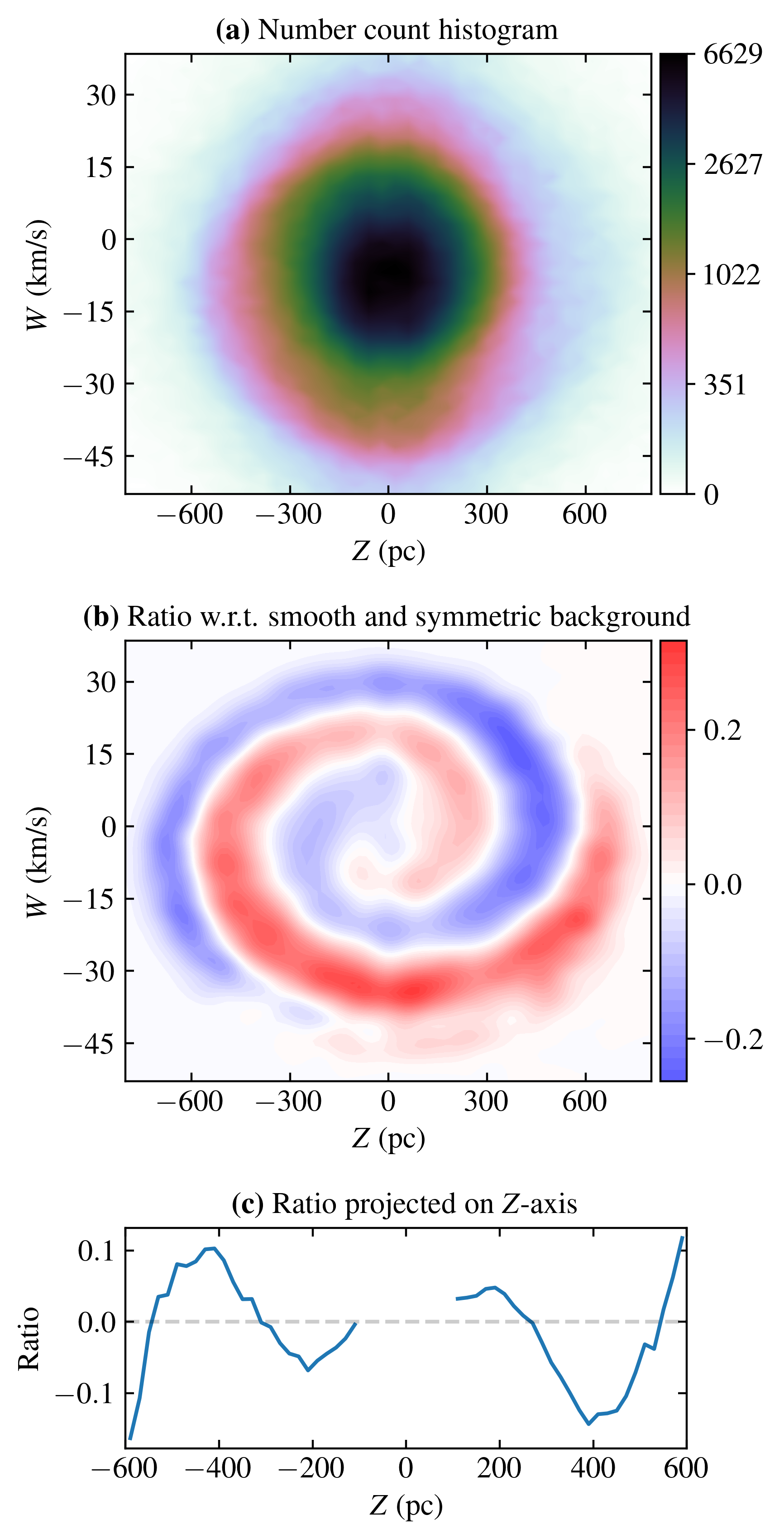}
    \caption{
    Phase-space spiral of the immediate Solar neighbourhood ($R_\text{cyl}<500~\pc$). The three panels show: \textbf{(a)} the stellar number count density in the $(Z,W)$ phase-space plane; \textbf{(b)} a ratio of this histogram with respect to a best-fit smooth and symmetric background distribution; \textbf{(c)} a ratio with respect to the same background distribution, but projected on the $Z$-axis. The over-densities at $Z \simeq 200~\pc$ and $Z \simeq -400~\pc$ have clear counterparts in Fig.~\ref{fig:Rz_residuals_2<G<3}. In panel \textbf{(b)}, we exclude regions far from the panel centre, where the total number count is low and the statistical noise is high. In panel \textbf{(c)}, we mask $|Z|<100~\pc$, where the \emph{Gaia} radial velocity sample is dominated by strong selection effects due to stellar crowding. Further details are found in the text.
    }
    \label{fig:spiral_proj}
\end{figure}

In Figure~\ref{fig:Rz_residuals_2<G<3}, we show the stellar number density for the $2<M_G\leq 3$ data sample in the $(R,Z)$-plane, for the region $|Y|<250~\pc$. We also show its ratio with respect to $f_\text{symm.}$. We clearly see the projection of the phase-space spiral, which appears as over-densities at $Z\simeq 250~\pc$ and $Z\simeq -400~\pc$ for $R$ in range of roughly 7--9.5~kpc. This connection is illustrated in Fig.~\ref{fig:spiral_proj}, where we show the phase-space spiral of the Solar neighbourhood, for the spatial cylindrical volume that fulfils $R_\text{cyl} \equiv \sqrt{X^2+Y^2} <500~\pc$, and stars with available radial velocity measurements. These results come from \cite{2021A&A...653A..86W}; we refer to that article for a detailed explanation of the method and data quality cuts. The top panel of Fig.~\ref{fig:spiral_proj} shows the stellar number count histogram in the $(Z,W)$-plane, while the middle panel shows how this stellar number density compares to a fitted smooth and symmetric background distribution. This background distribution is a Gaussian mixture model, consisting of six Gaussians that are all constrained to be centred on the same point in the $(Z,W)$-plane. The bottom panel shows how the spiral perturbation is projected onto the vertical spatial axis (i.e. how it manifests in terms of an $n(Z)$ perturbation). In the immediate Solar neighbourhood, it corresponds to over-densities at $Z \simeq 200~\pc$ and $Z \simeq -400~\pc$, and under-densities at the corresponding $Z \simeq -200~\pc$ and $Z \simeq 400~\pc$, which is clearly consistent with the results shown in Figure~\ref{fig:Rz_residuals_2<G<3}. It is difficult to tell whether these structures continue outside this range in $R$. Moreover, it is unclear whether our results can be trusted at such great distances, especially so close to the disk mid-plane. The large-scale asymmetry seen at greater heights in Figure~\ref{fig:disk_residuals_2<G<3} is also evident in both panels of Figure~\ref{fig:Rz_residuals_2<G<3}. The figure suggests that there is a misalignment between between the stellar populations occupying large heights above and below the plane ($|Z|\simeq 600~\pc$) and those at smaller heights ($\leq 300~\pc$), with the mid-plane of the former population exhibiting a positive slope with respect to $R$, while the latter population is flat, especially for $R > 9~\kpc$.

Our discussion of number densities in this section has focused on the data sample defined by $2 < M_G \leq 3$, which we consider to be most informative. The brighter data samples reach greater distances but it is more difficult to tease out clear stellar number density structures since the information gathered at those distances is plagued by poorer statistics and systematic issues that we have not been able to control for (e.g. degeneracies between dust extinction and distance). Conversely, the dimmest data sample has greatest number of stars and yields robust and trustworthy results, but also covers a smaller spatial volume. The corresponding plots of these other data samples can be found in Appendix~\ref{sec:other_plots}. Overall, similar stellar number density structures are visible in all four stellar samples, although the perturbations at lower vertical energies are less pronounced for the dimmest data sample.

\begin{figure*}
	\centering
	\includegraphics[width=0.96\textwidth]{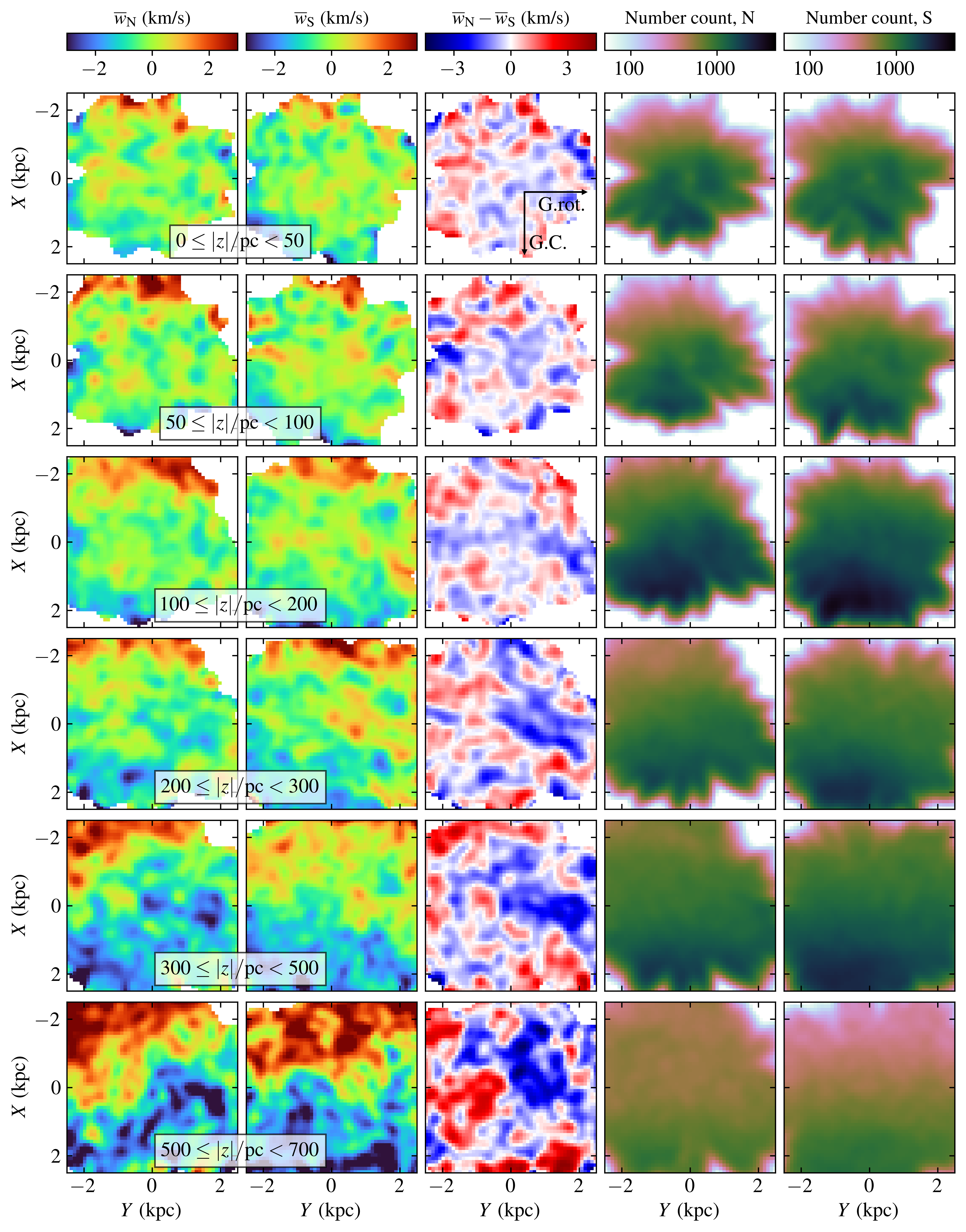}
    \caption{Mean vertical velocities of the data sample with $0 < M_G \leq 1$, in the same spatial volumes as in Fig.~\ref{fig:disk_residuals_2<G<3}. The results of each bin in $z$ are averaged over a larger area in the $(X,Y)$-plane for better visibility; the $(X,Y)$-grid is convolved with a 2d Gaussian with a standard deviation of 150~pc. The two right-hand columns show the number count in the respective spatial volumes; these quantities account for the smoothing in the $(X,Y)$-plane and correspond to the effective number of stars that inform the $\overline{w}$ value. A volume cell is masked if this effective stellar number count falls below 100.}
    \label{fig:disk_W_0<G<1}
\end{figure*}

In Fig.~\ref{fig:disk_W_0<G<1}, we show the mean vertical velocity distribution for our brightest data sample in the same volume cells that were used in Fig.~\ref{fig:disk_residuals_2<G<3}. Due to a smaller amount of statistics for the velocity information, we smoothed these maps in the $(X,Y)$-plane by convolving it with a 2d Gaussian with a standard deviation of 150~pc in both directions, corresponding to an effective area of $1.4 \times 10^5~\pc^2$. The vertical velocity is offset by $7.25~\kmsec$ to account for the Sun's motion with respect to the mid-plane. The corresponding plots for two other data samples can be found in Appendix~\ref{sec:other_plots}, although they are much more limited in distance. The two main stellar number density perturbations that we saw in Fig.~\ref{fig:disk_residuals_2<G<3} have clear counterparts in the vertical velocity field. First, the over-density that is close to the Galactic mid-plane at approximately $(X,Y)=(-0.3,0.8)~\kpc$ has a vertical velocity counterpart with a similar shape. The feature is seen most clearly in the fourth row of Fig.~\ref{fig:disk_W_0<G<1}, with negative values for $\overline{w}_\text{N}-\overline{w}_\text{S}$ in the third column, implying compression. Second, the large scale asymmetries at greater heights have a corresponding structure in the vertical velocity field, as can be seen in the two bottom rows of Fig.~\ref{fig:disk_W_0<G<1}; towards the Galactic anti-centre, both the north and south have a positive mean vertical velocity. As with the stellar number density perturbation, the feature is present at larger distances from the mid-plane.

\begin{figure}
	\centering
	\includegraphics[width=1\columnwidth]{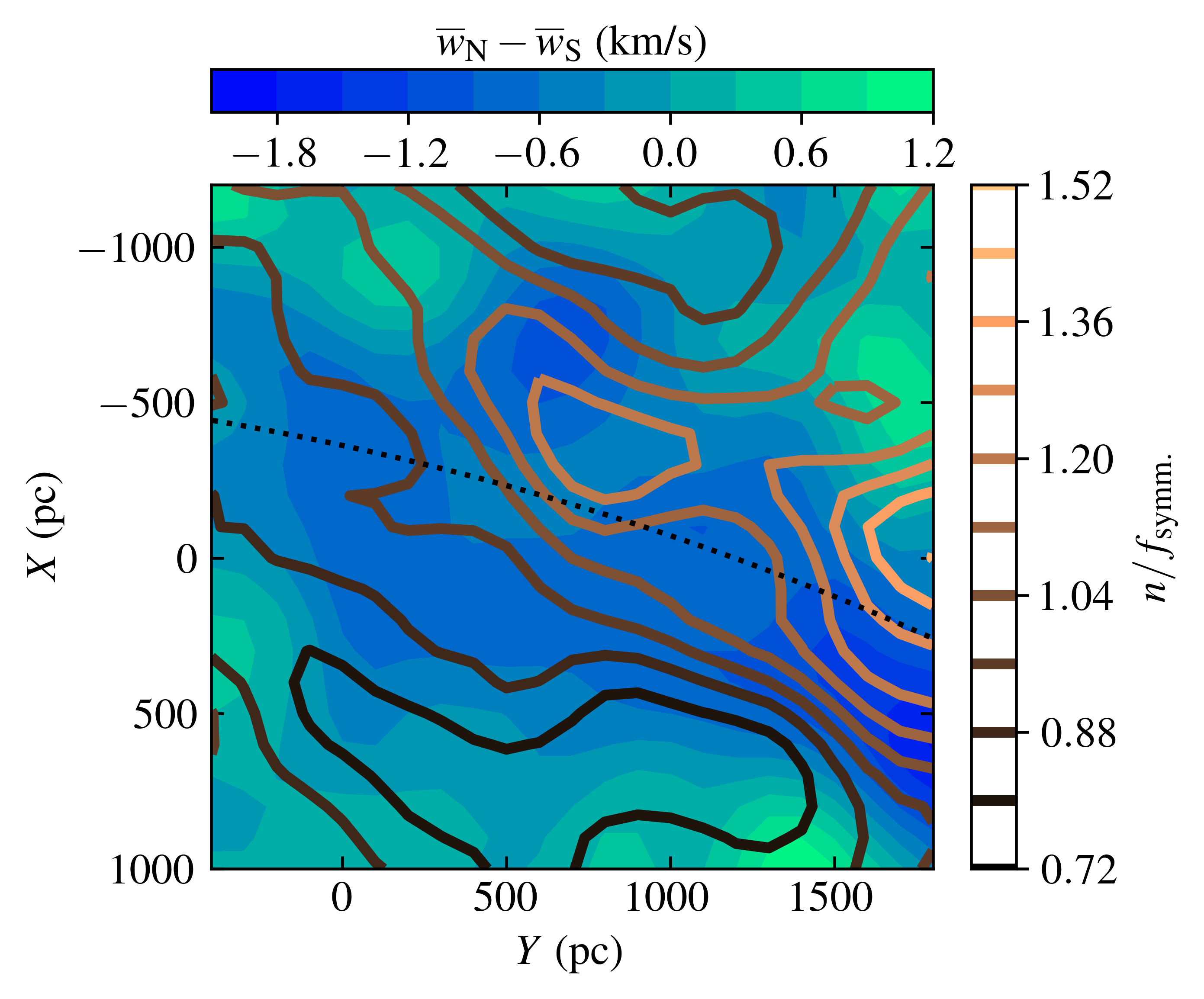}
    \caption{Joint stellar number density perturbation and vertical velocity perturbation in the disk plane, for the data sample with absolute magnitude in $0 < M_G \leq 1$, integrated over $|z|<300~\pc$. The mean vertical velocity distribution is smoothed over 150 pc in $X$ and $Y$ for better visibility. The dotted lines corresponds to the location of the Local Spiral Arm, according to \cite{Reid2014}.}
    \label{fig:nuWcomparison_0<G<1}
\end{figure}

\begin{figure}
	\centering
	\includegraphics[width=1\columnwidth]{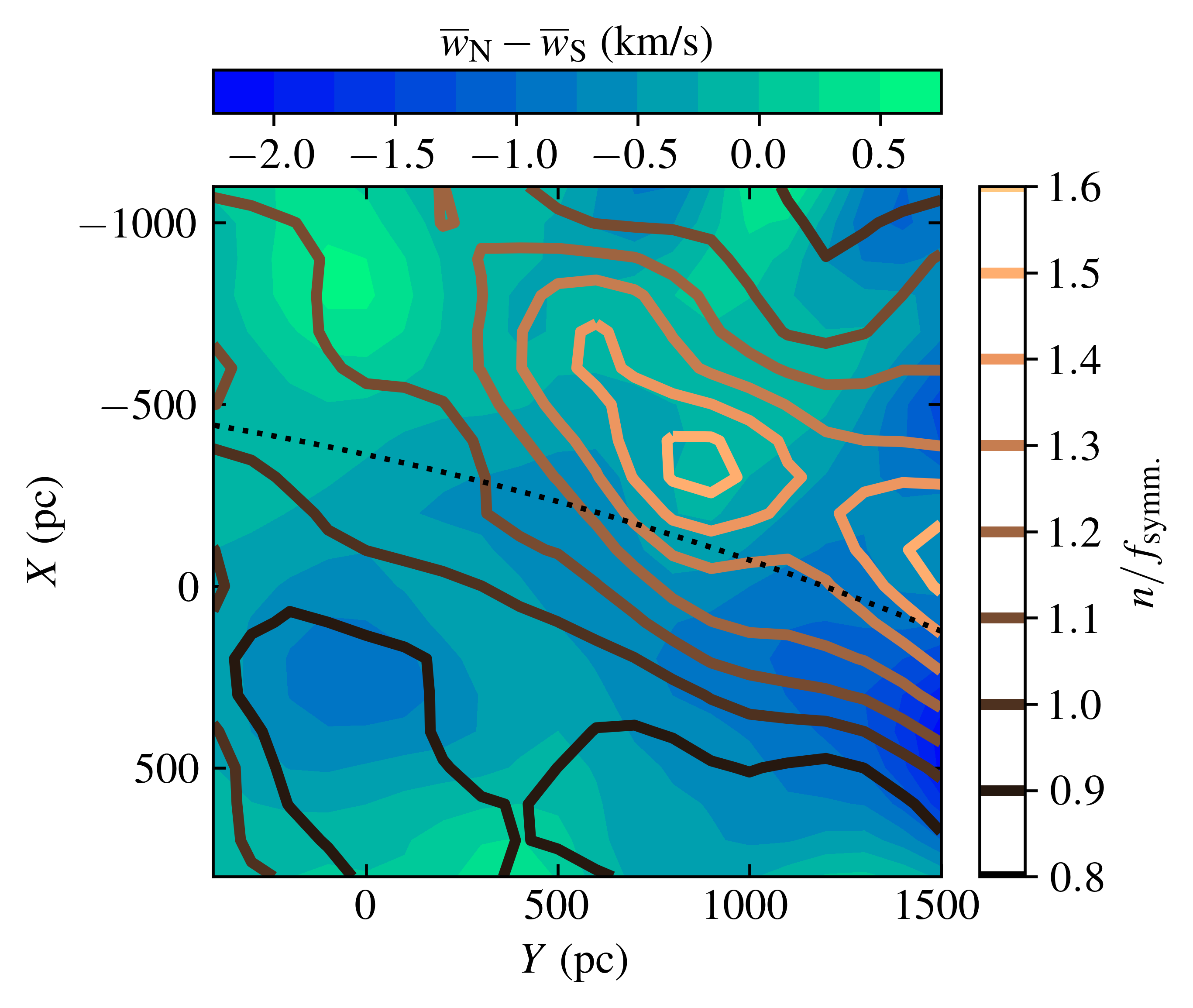}
    \caption{Same as Fig.~\ref{fig:nuWcomparison_0<G<1}, but for the data sample with absolute magnitude in $1 < M_G \leq 2$. The range in $X$ and $Y$ is slightly different in this figure, due to the distance limit imposed by $v_\text{RV}$ observations.}
    \label{fig:nuWcomparison_1<G<2}
\end{figure}

In Figs.~\ref{fig:nuWcomparison_0<G<1} and \ref{fig:nuWcomparison_1<G<2}, we show joint contour plots of the stellar number density and vertical velocity perturbations in the spatial region where we saw an elongated perturbation at lower heights, in both $n$ and $W$. The disk plane area covered in these two figures is determined by the distance limits of the respective data samples, mainly from the $\overline{W}$ field, which requires $v_\text{RV}$ measurements; for the same reason, the two dimmer data samples are too limited in distance to be informative of this spatial region. The figures highlight the relationship between the two fields. By eye, the perturbations in density and vertical velocity have roughly the same orientation and the same width across the short axis, but are out of phase by $\pi/2$. This general structure is present in both data samples and figures. Simultaneous measurements of a perturbation in $n$ and $W$ allow us to associated a timescale with the disturbance. The continuity equation can be written
\begin{equation}
\frac{1}{n}\frac{dn}{dt} = -\nabla\cdot {\bf V},
\end{equation}
which gives the timescale 
$\tau = \left (\delta n/n\right )/\left (\Delta W/\Delta z\right )$. The perturbation described here, $\Delta W\simeq 3\,\kmsec$ for $\Delta z\simeq 600~\pc$ and $\delta n/n\simeq 0.4$, which gives $\tau \simeq 80~\Myr$. We note that this calculation neglects a stellar source term, which could be significant for the star forming region of a spiral arm, especially for more luminous stars; this is discussed further in Sect.~\ref{sec:discussion}. For a further discussion of the divergence of the local stellar velocity field, see \citet{monari2015} and \citet{nelson2022}.

\section{Discussion}\label{sec:discussion}

Evidently, the stellar number density and vertical velocity fields show evidence for perturbations across our local patch of the disk. The features that we associate with perturbations are present in all four samples, though they differ in amplitude and structure from one sample to the next. For example, the elongated feature centred on $(X,Y) \simeq (-0.3,0.8)~\kpc$ and at heights $|z|\lesssim 300~\pc$ is most prominent in the brighter samples. We have tentatively identified this feature with the Local Spiral Arm and so the magnitude dependence of the feature may reflect the observation that spiral structure is associated with recent star formation and hence stars at the bright end of the luminosity function \citep{binney1998}. The general statement is that perturbations in stellar number density $n$ do not perfectly reflect those of the total stellar mass density, though they are clearly related.

To explore the possible connection of this feature with the Local Spiral Arm \citep{Xu2013}, we zoom into this region in Figs.~\ref{fig:nuWcomparison_0<G<1} and \ref{fig:nuWcomparison_1<G<2}. We see that the $n$ and $\overline{W}$ perturbations are offset by approximately a quarter wavelength. This suggests a breathing wave that is travelling in the direction of $l\simeq 270$~deg, roughly coincident with the position of the Local Spiral Arm found by \citet{Reid2014}. The link between spiral structure and breathing modes has been established in an analytic study of the linearized Boltzmann equation by \citet{monari2016a} and in an analysis of high-resolution simulation by \citet{Kumar2022}. In a related work, \citet{monari2016b} showed that a strong Galactic bar can alter and repress the phase offset between the $n$ and $\overline{W}$ perturbations; this scenario is disfavoured by our results.

The following toy model illustrates the breathing-mode hypothesis. For simplicity, we assume that the local gravitational potential is additively separable in $R$ and $z$ and that the vertical component of the potential is harmonic with $\Phi(z) = \frac{1}{2}\Omega_z^2 z^2$. The vertical action-angle variables are then $J_z = E_z/\Omega_z$ and $\theta_z = \tan^{-1}(\Omega_z z/w)$ where $E_z = w^2/2 + \Phi_z(z)$ is the vertical energy. Vertical oscillations follow a clockwise path (that is, increasing $\theta_z$) in the $(z,w)$-plane. For definiteness, we imagine that the unperturbed system is isothermal in the vertical direction so that the equilibrium DF in the $(z,w)$-plane is $f_0 \propto \exp{(-E_z/\sigma_z^2)}$. The simplest breathing mode perturbation is proportional to $\cos(2\theta_z - \omega_b t)$. At $t=0$, the DF is squeezed in $z$ and stretched in $w$, thereby increasing the density near the mid-plane. The perturbed DF then rotates in the clockwise sense with a pattern speed $\omega_b/2$. The complete model is 
\begin{equation}
    f({\bf X},W) = f_0(E_z) \{1 + \epsilon E_z\cos [
    2\theta_z - \omega_b t - \chi(X,Y) ] \},
\end{equation}
where $\chi$ encodes the propagation of the wave in the plane of the disk. Though this model is purely phenomenological, its functional form is motivated by analytic studies of modes in an isothermal plane \citep{mathur1990,weinberg1991,widrow2015}.

\begin{figure}
	\centering
	\includegraphics[width=1\columnwidth]{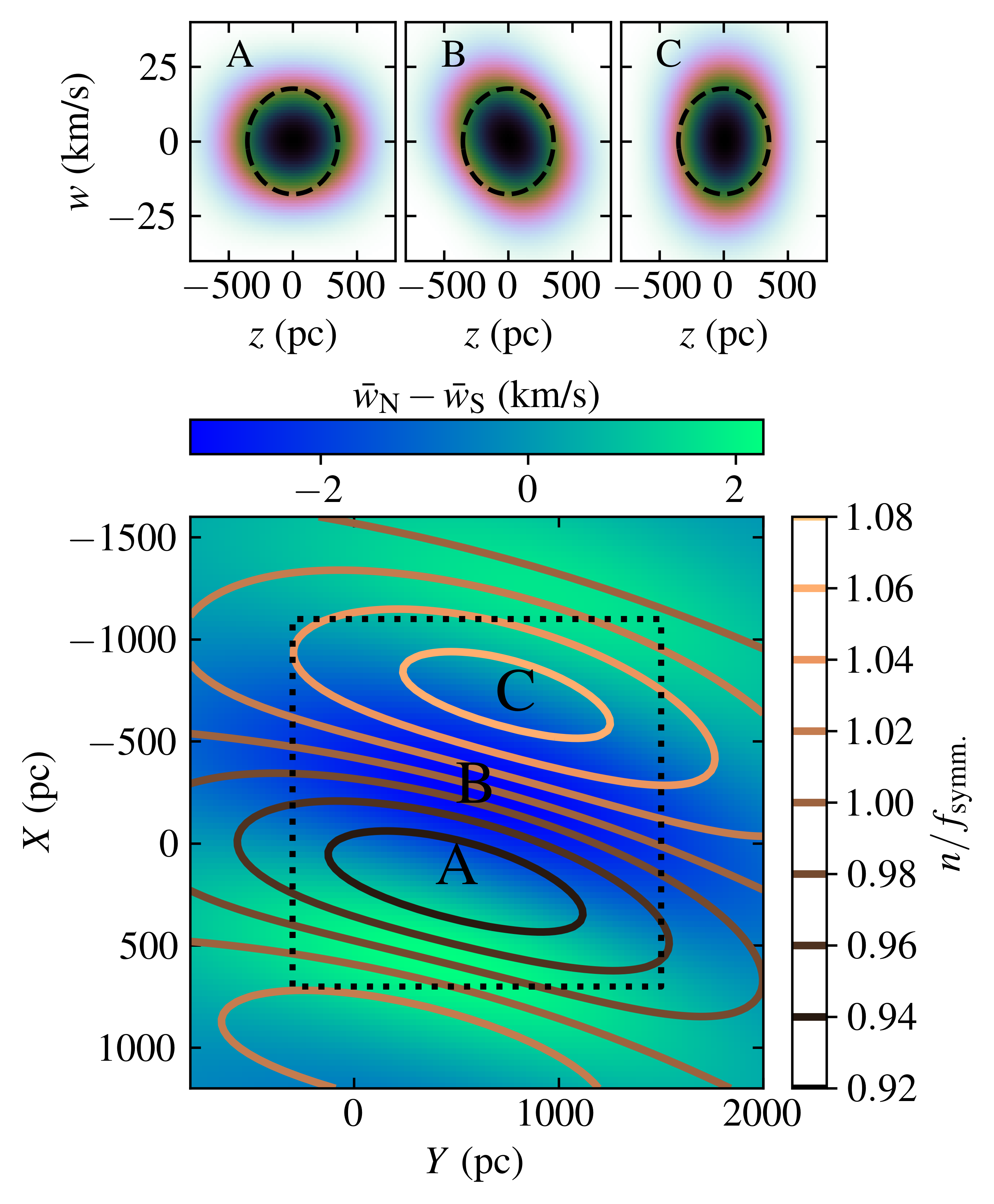}
    \caption{Toy model perturbations to the disk. The bottom panel shows $n$ and $\overline{W}$ for $|z|<300~\pc$ and is analogous to Figs.~\ref{fig:nuWcomparison_0<G<1} and \ref{fig:nuWcomparison_1<G<2}. The dotted line corresponds to the disk area covered in Fig.~\ref{fig:nuWcomparison_0<G<1}. In the top three panels, we show the number counts in the $(z,w)$-plane at the points A, B, and C that are highlighted in the lower panel. In order to facilitate a comparison between them, an iso-energy contour for the unperturbed disk is shown as a dashed line.}
    \label{fig:toymodel}
\end{figure}

In Fig.~\ref{fig:toymodel} we present a chi-by-eye realisation of the model that captures qualitative features of Figs.~\ref{fig:nuWcomparison_0<G<1} and \ref{fig:nuWcomparison_1<G<2}. The function $\chi$ is chosen to correspond to an outward propagating, trailing logarithmic spiral:
\begin{equation}
    \chi(X,Y) = k\log(R/R_0) - p\phi,
\end{equation}
where $R$ and $\phi$ are Galactocentric polar coordinates, $p$ is the tangent of the pitch angle, and the wavelength is $2\pi R_0/k$. For the figure, we set $k$ and $p$ so that the wavelength is $1\,{\rm kpc}$ and the pitch angle is $12^\circ$, as is the case for the Local Spiral Arm \citep{Xu2013}. We have also included an envelope function that serves to localise the perturbation about the point $(X,Y) = (-200,\,600)~\pc$. The three top panels in Fig.~\ref{fig:toymodel} correspond to $\omega_b t = \{\pi, \, 3\pi/2, \, 2\pi\}$, respectively.

The density and velocity fields of this simple toy model capture the qualitative features seen in the data. However, as discussed in the previous section, the data exhibit a number density perturbation of order $40~\%$ and a velocity perturbation of a few $\kmsec$. Our simple toy model predicts stronger perturbation for the velocity field relative to the number density field (where the overall strength is set by the free parameter $\epsilon$). Evidently, an explanation of the observed relative strength of the velocity and density perturbations will require a more complicated model. An obvious extension would be to consider a superposition of modes. Furthermore, as mentioned in the beginning of this section, the observed perturbation in $n$ is likely affected by recent star formation, especially for our brighter data samples. Hence, the over-density in $n$ is likely inflated as compared to the relative over-density of the total matter density field in the same spatial location.

The large scale bending mode feature is seen as a upward shift in the thicker disk component (roughly $|z|>300~\pc$), for stars in the direction of the Galactic anti-centre. The same structure is reflected in the vertical velocity distribution, where the corresponding northern and southern spatial volumes have a mean velocity towards the Galactic north, thus having the characteristic of a bending mode. However, the thinner disk component is less affected within the studied range in distance, remaining much more flat for both $n$ and $\overline{W}$. The structure can be interpreted as a mix of smaller-scale bending waves and the global Galactic warp as it extends into the Solar neighbourhood. This interpretation is consistent with the analysis by \cite{schonrich2018}, who measured $\overline{W}$ as a function of $L_z$ (the angular momentum about the $z$-axis) for the immediate Solar neighbourhood stars in the \emph{Gaia}-TGAS dataset. They found that the variations in $\overline{W}$ could be modelled as small-scale oscillations of $\overline{W}$ with $R$ (with a wavelength of roughly 2.5~kpc) superimposed on a linear function that increases with $R$. This linear trend is also consistent with the results from \cite{Poggio2020}, who modelled the large-scale Galactic warp and precession of the Milky Way's stellar disk. In their model, even though the Sun lies just 17~deg from the line-of-nodes, constant $W$ contours were roughly aligned with Galactic azimuth. We find a similar alignment, as seen in Fig.~\ref{fig:disk_W_0<G<1}. The situation with $n$ is more complicated. In the model by \cite{Poggio2020}, the disk bends toward the south (negative $z$) in the Solar neighbourhood. Thus, one expects that the north--south asymmetry in $n$ should increase with $R$. Due to our close proximity with the line-of-nodes, the direction of steepest increase in the asymmetry will be in the direction of increasing $R$ and decreasing $\phi$ (increasing $X$ and decreasing $Y$.). This trend is consistent with what we found in Fig.~\ref{fig:disk_residuals_2<G<3} for the regions closest to the mid-plane. However, the sign reverses for greater $|z|$. This complicated structure in number density can also be seen in Fig.~\ref{fig:Rz_residuals_2<G<3}.

Some structures in $n$, such as the horizontal bands in the bottom panel of Fig.~\ref{fig:Rz_residuals_2<G<3}, with $|Z|$ in range 200--600~pc, are projections of the phase-space spiral. The properties of the phase space spiral such as its phase and amount of winding in the $(Z,W)$-plane, vary slowly across the disk on scales of a few kiloparsecs in $X$ and $Y$ (e.g. \citealt{Bland-Hawthorn2019,2021A&A...653A..86W,widmark2021_spiralsIII,Hunt2022}; also supported by simulations, e.g. \citealt{2021MNRAS.508.1459H}). In Appendix~\ref{app:spiral_plots}, we show how the spiral angle in the $(Z,W)$-plane vary with $X$ and $Y$, which in turn translates into how the spiral perturbation projects onto the $Z$-axis. At lower heights ($Z\simeq 200~\pc$), we see a clear correspondence between the spiral density perturbation's projection in $n(Z)$ and the north-south asymmetries seen in the fourth and fifth panel rows of Fig.~\ref{fig:disk_residuals_2<G<3}. Conversely, the two main density perturbations that we have identified in this work, a small-scale breathing mode that we tentatively associate with the Local Spiral Arm and a large-scale bending mode, do not match the properties of a projected phase-space spiral; the former is much too localised in space and is symmetric; the latter is a very large relative perturbation found mainly at greater heights and does not match the azimuthal variation of the phase-space spiral in this spatial region. We refer to Appendix~\ref{app:spiral_plots} for further details.

There are likely systematic effects that bias our results, especially at greater distances and in the general direction of the Galactic centre, where dust extinction and stellar crowding are more severe. In principle, there could be some confounding systematic that creates a spatially dependent distance bias, for example arising from dust clouds, which could affect both the $n$ and $\overline{W}$ fields in the same spatial region. However, this is not likely to explain the two main perturbations that we have identified in this work. For the Local Spiral Arm, the structure in $n$ and $\overline{W}$ is elongated and close to the solar position, such that the viewing angle relative to its axis of elongation varies significantly. Despite this, we see a qualitatively similar structure over its axis of elongation. For the large-scale bending mode, its presence at greater heights makes it much less affected by stellar crowding and dust extinction, and we also see it over a large portion of the sky. Furthermore, we see both of these structures in all data samples, at least to the extent that they probe those spatial volumes.

\section{Conclusion}\label{sec:conclusion}

In this work, we have mapped the stellar number density distribution ($n$) and the mean vertical velocity distribution ($\overline{W}$), as a function of spatial position in the Milky Way disk, out to a distance of a few kiloparsecs. We have done so in a fairly model independent manner using GPs, which does not rely on any symmetry assumptions.

Apart from projections of the phase-space spiral, we identify two main perturbation features with respect to a fully symmetric background. First, we see an elongated over-density feature in $n$ and corresponding breathing mode compression in $\overline{W}$ at the spatial location of the Local Spiral Arm. The ridges of these $n$ and $\overline{W}$ structures are offset in the direction perpendicular to the spiral arm, indicating a travelling breathing mode. Second, we see a large-scale bending mode feature in both $n$ and $\overline{W}$. We make the novel observation that within our studied spatial volume, out to a distance of at least 2~kpc in the direction of the Galactic anti-centre, this bending mode feature affects the stellar number density at greater heights, while the thinner disk component ($|z|\lesssim 300~\pc$) remains more flat in both $n$ and $\overline{W}$.

An obvious extension of this work would be to combine a smooth model for the number density field with a model for the full three-dimensional velocity field. This would allow one to use the continuity and Jeans equations to more fully explore the connections between vertical motions and spiral arms as well as other examples of disequilibrium in the disk \citep{monari2015, monari2016a, monari2016b, nelson2022}.

We have demonstrated that with a careful treatment of selection effects, the stellar number density distribution can be mapped, even in fairly distant regions of the thin stellar disk. With more sophisticated and accurate complementary distance estimations, using photometric or spectroscopic information, in synergy with improved three-dimensional dust maps, we expect to reach even greater distances and depths in the near future.

\begin{acknowledgements}

We would like to thank Friedrich Anders and Giacomo Monari for useful discussions. We also want to thank the anonymous referee for a thorough and constructive report. AW acknowledges support from the Carlsberg Foundation via a Semper Ardens grant (CF15-0384). APN is supported by a Research Leadership Award from the Leverhulme Trust.
LMW acknowledges the financial support of the Natural Sciences and Engineering Research Council of Canada. This work made use of an HPC facility funded by a grant from VILLUM FONDEN (projectnumber 16599).

This work has made use of data from the European Space Agency (ESA) mission \emph{Gaia} (\url{https://www.cosmos.esa.int/gaia}), processed by the \emph{Gaia} Data Processing and Analysis Consortium (DPAC,
\url{https://www.cosmos.esa.int/web/gaia/dpac/consortium}). Funding for the DPAC has been provided by national institutions, in particular the institutions participating in the \emph{Gaia} Multilateral Agreement.

This research utilised the following open-source Python packages: \textsc{Matplotlib} \citep{matplotlib}, \textsc{NumPy} \citep{numpy}, \textsc{George} \citep{georgeGP}, \textsc{healpy} \citep{healpy}.

\end{acknowledgements}

\bibliographystyle{aa} 
\bibliography{main} 

\begin{thebibliography}{55}
\expandafter\ifx\csname natexlab\endcsname\relax\def\natexlab#1{#1}\fi

\bibitem[{{Ambikasaran} {et~al.}(2015){Ambikasaran}, {Foreman-Mackey},
  {Greengard}, {Hogg}, \& {O'Neil}}]{georgeGP}
{Ambikasaran}, S., {Foreman-Mackey}, D., {Greengard}, L., {Hogg}, D.~W., \&
  {O'Neil}, M. 2015, IEEE Transactions on Pattern Analysis and Machine
  Intelligence, 38, 252

\bibitem[{{Anders} {et~al.}(2022){Anders}, {Khalatyan}, {Queiroz}, {Chiappini},
  {Ard{\`e}vol}, {Casamiquela}, {Figueras}, {Jim{\'e}nez-Arranz}, {Jordi},
  {Mongui{\'o}}, {Romero-G{\'o}mez}, {Altamirano}, {Antoja}, {Assaad},
  {Cantat-Gaudin}, {Castro-Ginard}, {Enke}, {Girardi}, {Guiglion}, {Khan},
  {Luri}, {Miglio}, {Minchev}, {Ramos}, {Santiago}, \&
  {Steinmetz}}]{2022A&A...658A..91A}
{Anders}, F., {Khalatyan}, A., {Queiroz}, A.~B.~A., {et~al.} 2022, A\&A, 658,
  A91

\bibitem[{{Antoja} {et~al.}(2018){Antoja}, {Helmi}, {Romero-G{\'o}mez}, {Katz},
  {Babusiaux}, {Drimmel}, {Evans}, {Figueras}, {Poggio}, {Reyl{\'e}}, {Robin},
  {Seabroke}, \& {Soubiran}}]{antoja2018}
{Antoja}, T., {Helmi}, A., {Romero-G{\'o}mez}, M., {et~al.} 2018, \nat, 561,
  360

\bibitem[{{Bennett} \& {Bovy}(2019)}]{bennett2019}
{Bennett}, M. \& {Bovy}, J. 2019, \mnras, 482, 1417

\bibitem[{{Binney} \& {Merrifield}(1998)}]{binney1998}
{Binney}, J. \& {Merrifield}, M. 1998, {Galactic Astronomy}

\bibitem[{{Bland-Hawthorn} {et~al.}(2019){Bland-Hawthorn}, {Sharma},
  {Tepper-Garcia}, {Binney}, {Freeman}, {Hayden}, {Kos}, {De Silva}, {Ellis},
  {Lewis}, {Asplund}, {Buder}, {Casey}, {D'Orazi}, {Duong}, {Khanna}, {Lin},
  {Lind}, {Martell}, {Ness}, {Simpson}, {Zucker}, {Zwitter}, {Kafle},
  {Quillen}, {Ting}, \& {Wyse}}]{Bland-Hawthorn2019}
{Bland-Hawthorn}, J., {Sharma}, S., {Tepper-Garcia}, T., {et~al.} 2019, \mnras,
  486, 1167

\bibitem[{{Boubert} \& {Everall}(2020)}]{Boubert-II}
{Boubert}, D. \& {Everall}, A. 2020, \mnras, 497, 4246

\bibitem[{{Cantat-Gaudin} {et~al.}(2022){Cantat-Gaudin}, {Fouesneau}, {Rix},
  {Brown}, {Castro-Ginard}, {Drimmel}, {Hogg}, {Casey}, {Khanna}, {Oh}, {Price
  Whelan}, {Belokurov}, {Saydjari}, \& {Green}}]{CantatGaudin2022}
{Cantat-Gaudin}, T., {Fouesneau}, M., {Rix}, H.-W., {et~al.} 2022, arXiv
  e-prints, arXiv:2208.09335

\bibitem[{{Cantat-Gaudin} {et~al.}(2018){Cantat-Gaudin}, {Jordi}, {Vallenari},
  {Bragaglia}, {Balaguer-N{\'u}{\~n}ez}, {Soubiran}, {Bossini}, {Moitinho},
  {Castro-Ginard}, {Krone-Martins}, {Casamiquela}, {Sordo}, \&
  {Carrera}}]{2018A&A...618A..93C}
{Cantat-Gaudin}, T., {Jordi}, C., {Vallenari}, A., {et~al.} 2018, \aap, 618,
  A93

\bibitem[{{Carlin} {et~al.}(2013){Carlin}, {DeLaunay}, {Newberg}, {Deng},
  {Gole}, {Grabowski}, {Jin}, {Liu}, {Liu}, {Luo}, {Yuan}, {Zhang}, {Zhao}, \&
  {Zhao}}]{carlin2013}
{Carlin}, J.~L., {DeLaunay}, J., {Newberg}, H.~J., {et~al.} 2013, \apjl, 777,
  L5

\bibitem[{{Everall} {et~al.}(2022{\natexlab{a}}){Everall}, {Belokurov},
  {Evans}, {Boubert}, \& {Grand}}]{everall-II}
{Everall}, A., {Belokurov}, V., {Evans}, N.~W., {Boubert}, D., \& {Grand}, R.
  J.~J. 2022{\natexlab{a}}, \mnras, 511, 3863

\bibitem[{{Everall} \& {Boubert}(2022{\natexlab{a}})}]{Everall-V}
{Everall}, A. \& {Boubert}, D. 2022{\natexlab{a}}, \mnras, 509, 6205

\bibitem[{{Everall} \& {Boubert}(2022{\natexlab{b}})}]{everall2022}
{Everall}, A. \& {Boubert}, D. 2022{\natexlab{b}}, \mnras, 509, 6205

\bibitem[{{Everall} {et~al.}(2022{\natexlab{b}}){Everall}, {Evans},
  {Belokurov}, {Boubert}, \& {Grand}}]{everall-I}
{Everall}, A., {Evans}, N.~W., {Belokurov}, V., {Boubert}, D., \& {Grand}, R.
  J.~J. 2022{\natexlab{b}}, \mnras, 511, 2390

\bibitem[{{Friske} \& {Sch{\"o}nrich}(2019)}]{friske2019}
{Friske}, J. K.~S. \& {Sch{\"o}nrich}, R. 2019, \mnras, 490, 5414

\bibitem[{{Gaia Collaboration} {et~al.}(2022{\natexlab{a}}){Gaia
  Collaboration}, {Drimmel}, {Romero-Gomez}, {Chemin}, {Ramos}, {Poggio},
  {Ripepi}, {Andrae}, {Blomme}, {Cantat-Gaudin}, {Castro-Ginard}, {Clementini},
  {Figueras}, {Fouesneau}, {Fremat}, {Jardine}, {Khanna}, {Lobel}, {Marshall},
  \& {Muraveva}}]{GaiaAsymmetry2022}
{Gaia Collaboration}, {Drimmel}, R., {Romero-Gomez}, M., {et~al.}
  2022{\natexlab{a}}, arXiv e-prints, arXiv:2206.06207

\bibitem[{{Gaia Collaboration} {et~al.}(2018){Gaia Collaboration}, {Katz},
  {Antoja}, {Romero-G{\'o}mez}, {Drimmel}, {Reyl{\'e}}, {Seabroke}, {Soubiran},
  {Babusiaux}, {Di Matteo}, {Figueras}, {Poggio}, {Robin}, {Evans}, {Brown},
  {Vallenari}, {Prusti}, {de Bruijne}, {Bailer-Jones}, {Biermann}, {Eyer},
  {Jansen}, {Jordi}, {Klioner}, {Lammers}, {Lindegren}, {Luri}, {Mignard},
  {Panem}, {Pourbaix}, {Randich}, {Sartoretti}, {Siddiqui}, {van Leeuwen},
  {Walton}, {Arenou}, {Bastian}, {Cropper}, {Lattanzi}, {Bakker}, {Cacciari},
  {Casta n}, {Chaoul}, {Cheek}, {De Angeli}, {Fabricius}, {Guerra}, {Holl},
  {Masana}, {Messineo}, {Mowlavi}, {Nienartowicz}, {Panuzzo}, {Portell},
  {Riello}, {Tanga}, {Th{\'e}venin}, {Gracia-Abril}, {Comoretto},
  {Garcia-Reinaldos}, {Teyssier}, {Altmann}, {Andrae}, {Audard},
  {Bellas-Velidis}, {Benson}, {Berthier}, {Blomme}, {Burgess}, {Busso},
  {Carry}, {Cellino}, {Clementini}, {Clotet}, {Creevey}, {Davidson}, {De
  Ridder}, {Delchambre}, {Dell'Oro}, {Ducourant},
  {Fern{\'a}ndez-Hern{\'a}ndez}, {Fouesneau}, {Fr{\'e}mat}, {Galluccio},
  {Garc{\'\i}a-Torres}, {Gonz{\'a}lez-N{\'u}{\~n}ez}, {Gonz{\'a}lez-Vidal},
  {Gosset}, {Guy}, {Halbwachs}, {Hambly}, {Harrison}, {Hern{\'a}ndez},
  {Hestroffer}, {Hodgkin}, {Hutton}, {Jasniewicz}, {Jean-Antoine-Piccolo},
  {Jordan}, {Korn}, {Krone-Martins}, {Lanzafame}, {Lebzelter}, {L{\"o}ffler},
  {Manteiga}, {Marrese}, {Mart{\'\i}n-Fleitas}, {Moitinho}, {Mora}, {Muinonen},
  {Osinde}, {Pancino}, {Pauwels}, {Petit}, {Recio-Blanco}, {Richards},
  {Rimoldini}, {Sarro}, {Siopis}, {Smith}, {Sozzetti}, {S{\"u}veges}, {Torra},
  {van Reeven}, {Abbas}, {Abreu Aramburu}, {Accart}, {Aerts}, {Altavilla},
  {{\'A}lvarez}, {Alvarez}, {Alves}, {Anderson}, {Andrei}, {Anglada Varela},
  {Antiche}, {Arcay}, {Astraatmadja}, {Bach}, {Baker},
  {Balaguer-N{\'u}{\~n}ez}, {Balm}, {Barache}, {Barata}, {Barbato}, {Barblan},
  {Barklem}, {Barrado}, {Barros}, {Barstow}, {Bartholom{\'e} Mu{\~n}oz},
  {Bassilana}, {Becciani}, {Bellazzini}, {Berihuete}, {Bertone}, {Bianchi},
  {Bienaym{\'e}}, {Blanco-Cuaresma}, {Boch}, {Boeche}, {Bombrun}, {Borrachero},
  {Bossini}, {Bouquillon}, {Bourda}, {Bragaglia}, {Bramante}, {Breddels},
  {Bressan}, {Brouillet}, {Br{\"u}semeister}, {Brugaletta}, {Bucciarelli},
  {Burlacu}, {Busonero}, {Butkevich}, {Buzzi}, {Caffau}, {Cancelliere},
  {Cannizzaro}, {Cantat-Gaudin}, {Carballo}, {Carlucci}, {Carrasco},
  {Casamiquela}, {Castellani}, {Castro-Ginard}, {Charlot}, {Chemin},
  {Chiavassa}, {Cocozza}, {Costigan}, {Cowell}, {Crifo}, {Crosta}, {Crowley},
  {Cuypers}, {Dafonte}, {Damerdji}, {Dapergolas}, {David}, {David}, {de
  Laverny}, {De Luise}, {De March}, {de Souza}, {de Torres}, {Debosscher}, {del
  Pozo}, {Delbo}, {Delgado}, {Delgado}, {Diakite}, {Diener}, {Distefano},
  {Dolding}, {Drazinos}, {Dur{\'a}n}, {Edvardsson}, {Enke}, {Eriksson},
  {Esquej}, {Eynard Bontemps}, {Fabre}, {Fabrizio}, {Faigler}, {Falc a},
  {Farr{\`a}s Casas}, {Federici}, {Fedorets}, {Fernique}, {Filippi},
  {Findeisen}, {Fonti}, {Fraile}, {Fraser}, {Fr{\'e}zouls}, {Gai}, {Galleti},
  {Garabato}, {Garc{\'\i}a-Sedano}, {Garofalo}, {Garralda}, {Gavel}, {Gavras},
  {Gerssen}, {Geyer}, {Giacobbe}, {Gilmore}, {Girona}, {Giuffrida}, {Glass},
  {Gomes}, {Granvik}, {Gueguen}, {Guerrier}, {Guiraud}, {Guti{\'e}}, {Haigron},
  {Hatzidimitriou}, {Hauser}, {Haywood}, {Heiter}, {Helmi}, {Heu}, {Hilger},
  {Hobbs}, {Hofmann}, {Holland}, {Huckle}, {Hypki}, {Icardi}, {Jan{\ss}en},
  {Jevardat de Fombelle}, {Jonker}, {Juh{\'a}sz}, {Julbe}, {Karampelas},
  {Kewley}, {Klar}, {Kochoska}, {Kohley}, {Kolenberg}, {Kontizas}, {Kontizas},
  {Koposov}, {Kordopatis}, {Kostrzewa-Rutkowska}, {Koubsky}, {Lambert},
  {Lanza}, {Lasne}, {Lavigne}, {Le Fustec}, {Le Poncin-Lafitte}, {Lebreton},
  {Leccia}, {Leclerc}, {Lecoeur-Taibi}, {Lenhardt}, {Leroux}, {Liao}, {Licata},
  {Lindstr{\o}m}, {Lister}, {Livanou}, {Lobel}, {L{\'o}pez}, {Managau}, {Mann},
  {Mantelet}, {Marchal}, {Marchant}, {Marconi}, {Marinoni}, {Marschalk{\'o}},
  {Marshall}, {Martino}, {Marton}, {Mary}, {Massari}, {Matijevi{\v{c}}},
  {Mazeh}, {McMillan}, {Messina}, {Michalik}, {Millar}, {Molina}, {Molinaro},
  {Moln{\'a}r}, {Montegriffo}, {Mor}, {Morbidelli}, {Morel}, {Morris},
  {Mulone}, {Muraveva}, {Musella}, {Nelemans}, {Nicastro}, {Noval},
  {O'Mullane}, {Ord{\'e}novic}, {Ord{\'o}{\~n}ez-Blanco}, {Osborne}, {Pagani},
  {Pagano}, {Pailler}, {Palacin}, {Palaversa}, {Panahi}, {Pawlak},
  {Piersimoni}, {Pineau}, {Plachy}, {Plum}, {Poujoulet}, {Pr{\v{s}}a},
  {Pulone}, {Racero}, {Ragaini}, {Rambaux}, {Ramos-Lerate}, {Regibo}, {Riclet},
  {Ripepi}, {Riva}, {Rivard}, {Rixon}, {Roegiers}, {Roelens}, {Rowell},
  {Royer}, {Ruiz-Dern}, {Sadowski}, {Sagrist{\`a} Sell{\'e}s}, {Sahlmann},
  {Salgado}, {Salguero}, {Sanna}, {Santana-Ros}, {Sarasso}, {Savietto},
  {Schultheis}, {Sciacca}, {Segol}, {Segovia}, {S{\'e}gransan}, {Shih},
  {Siltala}, {Silva}, {Smart}, {Smith}, {Solano}, {Solitro}, {Sordo}, {Soria
  Nieto}, {Souchay}, {Spagna}, {Spoto}, {Stampa}, {Steele},
  {Steidelm{\"u}ller}, {Stephenson}, {Stoev}, {Suess}, {Surdej}, {Szabados},
  {Szegedi-Elek}, {Tapiador}, {Taris}, {Tauran}, {Taylor}, {Teixeira},
  {Terrett}, {Teyssandier}, {Thuillot}, {Titarenko}, {Torra Clotet}, {Turon},
  {Ulla}, {Utrilla}, {Uzzi}, {Vaillant}, {Valentini}, {Valette}, {van Elteren},
  {Van Hemelryck}, {van Leeuwen}, {Vaschetto}, {Vecchiato}, {Veljanoski},
  {Viala}, {Vicente}, {Vogt}, {von Essen}, {Voss}, {Votruba}, {Voutsinas},
  {Walmsley}, {Weiler}, {Wertz}, {Wevers}, {Wyrzykowski}, {Yoldas},
  {{\v{Z}}erjal}, {Ziaeepour}, {Zorec}, {Zschocke}, {Zucker}, {Zurbach}, \&
  {Zwitter}}]{GaiaDR2Kinematics2018}
{Gaia Collaboration}, {Katz}, D., {Antoja}, T., {et~al.} 2018, \aap, 616, A11

\bibitem[{{Gaia Collaboration} {et~al.}(2016){Gaia Collaboration}, {Prusti},
  {de Bruijne}, {Brown}, {Vallenari}, {Babusiaux}, {Bailer-Jones}, {Bastian},
  {Biermann}, {Evans}, {Eyer}, {Jansen}, {Jordi}, {Klioner}, {Lammers},
  {Lindegren}, {Luri}, {Mignard}, {Milligan}, {Panem}, {Poinsignon},
  {Pourbaix}, {Randich}, {Sarri}, {Sartoretti}, {Siddiqui}, {Soubiran},
  {Valette}, {van Leeuwen}, {Walton}, {Aerts}, {Arenou}, {Cropper}, {Drimmel},
  {H{\o}g}, {Katz}, {Lattanzi}, {O'Mullane}, {Grebel}, {Holland}, {Huc},
  {Passot}, {Bramante}, {Cacciari}, {Casta{\~n}eda}, {Chaoul}, {Cheek}, {De
  Angeli}, {Fabricius}, {Guerra}, {Hern{\'a}ndez}, {Jean-Antoine-Piccolo},
  {Masana}, {Messineo}, {Mowlavi}, {Nienartowicz}, {Ord{\'o}{\~n}ez-Blanco},
  {Panuzzo}, {Portell}, {Richards}, {Riello}, {Seabroke}, {Tanga},
  {Th{\'e}venin}, {Torra}, {Els}, {Gracia-Abril}, {Comoretto},
  {Garcia-Reinaldos}, {Lock}, {Mercier}, {Altmann}, {Andrae}, {Astraatmadja},
  {Bellas-Velidis}, {Benson}, {Berthier}, {Blomme}, {Busso}, {Carry},
  {Cellino}, {Clementini}, {Cowell}, {Creevey}, {Cuypers}, {Davidson}, {De
  Ridder}, {de Torres}, {Delchambre}, {Dell'Oro}, {Ducourant}, {Fr{\'e}mat},
  {Garc{\'\i}a-Torres}, {Gosset}, {Halbwachs}, {Hambly}, {Harrison}, {Hauser},
  {Hestroffer}, {Hodgkin}, {Huckle}, {Hutton}, {Jasniewicz}, {Jordan},
  {Kontizas}, {Korn}, {Lanzafame}, {Manteiga}, {Moitinho}, {Muinonen},
  {Osinde}, {Pancino}, {Pauwels}, {Petit}, {Recio-Blanco}, {Robin}, {Sarro},
  {Siopis}, {Smith}, {Smith}, {Sozzetti}, {Thuillot}, {van Reeven}, {Viala},
  {Abbas}, {Abreu Aramburu}, {Accart}, {Aguado}, {Allan}, {Allasia},
  {Altavilla}, {{\'A}lvarez}, {Alves}, {Anderson}, {Andrei}, {Anglada Varela},
  {Antiche}, {Antoja}, {Ant{\'o}n}, {Arcay}, {Atzei}, {Ayache}, {Bach},
  {Baker}, {Balaguer-N{\'u}{\~n}ez}, {Barache}, {Barata}, {Barbier}, {Barblan},
  {Baroni}, {Barrado y Navascu{\'e}s}, {Barros}, {Barstow}, {Becciani},
  {Bellazzini}, {Bellei}, {Bello Garc{\'\i}a}, {Belokurov}, {Bendjoya},
  {Berihuete}, {Bianchi}, {Bienaym{\'e}}, {Billebaud}, {Blagorodnova},
  {Blanco-Cuaresma}, {Boch}, {Bombrun}, {Borrachero}, {Bouquillon}, {Bourda},
  {Bouy}, {Bragaglia}, {Breddels}, {Brouillet}, {Br{\"u}semeister},
  {Bucciarelli}, {Budnik}, {Burgess}, {Burgon}, {Burlacu}, {Busonero}, {Buzzi},
  {Caffau}, {Cambras}, {Campbell}, {Cancelliere}, {Cantat-Gaudin}, {Carlucci},
  {Carrasco}, {Castellani}, {Charlot}, {Charnas}, {Charvet}, {Chassat},
  {Chiavassa}, {Clotet}, {Cocozza}, {Collins}, {Collins}, {Costigan}, {Crifo},
  {Cross}, {Crosta}, {Crowley}, {Dafonte}, {Damerdji}, {Dapergolas}, {David},
  {David}, {De Cat}, {de Felice}, {de Laverny}, {De Luise}, {De March}, {de
  Martino}, {de Souza}, {Debosscher}, {del Pozo}, {Delbo}, {Delgado},
  {Delgado}, {di Marco}, {Di Matteo}, {Diakite}, {Distefano}, {Dolding}, {Dos
  Anjos}, {Drazinos}, {Dur{\'a}n}, {Dzigan}, {Ecale}, {Edvardsson}, {Enke},
  {Erdmann}, {Escolar}, {Espina}, {Evans}, {Eynard Bontemps}, {Fabre},
  {Fabrizio}, {Faigler}, {Falc{\~a}o}, {Farr{\`a}s Casas}, {Faye}, {Federici},
  {Fedorets}, {Fern{\'a}ndez-Hern{\'a}ndez}, {Fernique}, {Fienga}, {Figueras},
  {Filippi}, {Findeisen}, {Fonti}, {Fouesneau}, {Fraile}, {Fraser}, {Fuchs},
  {Furnell}, {Gai}, {Galleti}, {Galluccio}, {Garabato}, {Garc{\'\i}a-Sedano},
  {Gar{\'e}}, {Garofalo}, {Garralda}, {Gavras}, {Gerssen}, {Geyer}, {Gilmore},
  {Girona}, {Giuffrida}, {Gomes}, {Gonz{\'a}lez-Marcos},
  {Gonz{\'a}lez-N{\'u}{\~n}ez}, {Gonz{\'a}lez-Vidal}, {Granvik}, {Guerrier},
  {Guillout}, {Guiraud}, {G{\'u}rpide}, {Guti{\'e}rrez-S{\'a}nchez}, {Guy},
  {Haigron}, {Hatzidimitriou}, {Haywood}, {Heiter}, {Helmi}, {Hobbs},
  {Hofmann}, {Holl}, {Holland}, {Hunt}, {Hypki}, {Icardi}, {Irwin}, {Jevardat
  de Fombelle}, {Jofr{\'e}}, {Jonker}, {Jorissen}, {Julbe}, {Karampelas},
  {Kochoska}, {Kohley}, {Kolenberg}, {Kontizas}, {Koposov}, {Kordopatis},
  {Koubsky}, {Kowalczyk}, {Krone-Martins}, {Kudryashova}, {Kull}, {Bachchan},
  {Lacoste-Seris}, {Lanza}, {Lavigne}, {Le Poncin-Lafitte}, {Lebreton},
  {Lebzelter}, {Leccia}, {Leclerc}, {Lecoeur-Taibi}, {Lemaitre}, {Lenhardt},
  {Leroux}, {Liao}, {Licata}, {Lindstr{\o}m}, {Lister}, {Livanou}, {Lobel},
  {L{\"o}ffler}, {L{\'o}pez}, {Lopez-Lozano}, {Lorenz}, {Loureiro},
  {MacDonald}, {Magalh{\~a}es Fernandes}, {Managau}, {Mann}, {Mantelet},
  {Marchal}, {Marchant}, {Marconi}, {Marie}, {Marinoni}, {Marrese},
  {Marschalk{\'o}}, {Marshall}, {Mart{\'\i}n-Fleitas}, {Martino}, {Mary},
  {Matijevi{\v{c}}}, {Mazeh}, {McMillan}, {Messina}, {Mestre}, {Michalik},
  {Millar}, {Miranda}, {Molina}, {Molinaro}, {Molinaro}, {Moln{\'a}r},
  {Moniez}, {Montegriffo}, {Monteiro}, {Mor}, {Mora}, {Morbidelli}, {Morel},
  {Morgenthaler}, {Morley}, {Morris}, {Mulone}, {Muraveva}, {Musella},
  {Narbonne}, {Nelemans}, {Nicastro}, {Noval}, {Ord{\'e}novic},
  {Ordieres-Mer{\'e}}, {Osborne}, {Pagani}, {Pagano}, {Pailler}, {Palacin},
  {Palaversa}, {Parsons}, {Paulsen}, {Pecoraro}, {Pedrosa}, {Pentik{\"a}inen},
  {Pereira}, {Pichon}, {Piersimoni}, {Pineau}, {Plachy}, {Plum}, {Poujoulet},
  {Pr{\v{s}}a}, {Pulone}, {Ragaini}, {Rago}, {Rambaux}, {Ramos-Lerate},
  {Ranalli}, {Rauw}, {Read}, {Regibo}, {Renk}, {Reyl{\'e}}, {Ribeiro},
  {Rimoldini}, {Ripepi}, {Riva}, {Rixon}, {Roelens}, {Romero-G{\'o}mez},
  {Rowell}, {Royer}, {Rudolph}, {Ruiz-Dern}, {Sadowski}, {Sagrist{\`a}
  Sell{\'e}s}, {Sahlmann}, {Salgado}, {Salguero}, {Sarasso}, {Savietto},
  {Schnorhk}, {Schultheis}, {Sciacca}, {Segol}, {Segovia}, {Segransan},
  {Serpell}, {Shih}, {Smareglia}, {Smart}, {Smith}, {Solano}, {Solitro},
  {Sordo}, {Soria Nieto}, {Souchay}, {Spagna}, {Spoto}, {Stampa}, {Steele},
  {Steidelm{\"u}ller}, {Stephenson}, {Stoev}, {Suess}, {S{\"u}veges}, {Surdej},
  {Szabados}, {Szegedi-Elek}, {Tapiador}, {Taris}, {Tauran}, {Taylor},
  {Teixeira}, {Terrett}, {Tingley}, {Trager}, {Turon}, {Ulla}, {Utrilla},
  {Valentini}, {van Elteren}, {Van Hemelryck}, {van Leeuwen}, {Varadi},
  {Vecchiato}, {Veljanoski}, {Via}, {Vicente}, {Vogt}, {Voss}, {Votruba},
  {Voutsinas}, {Walmsley}, {Weiler}, {Weingrill}, {Werner}, {Wevers},
  {Whitehead}, {Wyrzykowski}, {Yoldas}, {{\v{Z}}erjal}, {Zucker}, {Zurbach},
  {Zwitter}, {Alecu}, {Allen}, {Allende Prieto}, {Amorim},
  {Anglada-Escud{\'e}}, {Arsenijevic}, {Azaz}, {Balm}, {Beck}, {Bernstein},
  {Bigot}, {Bijaoui}, {Blasco}, {Bonfigli}, {Bono}, {Boudreault}, {Bressan},
  {Brown}, {Brunet}, {Bunclark}, {Buonanno}, {Butkevich}, {Carret}, {Carrion},
  {Chemin}, {Ch{\'e}reau}, {Corcione}, {Darmigny}, {de Boer}, {de Teodoro}, {de
  Zeeuw}, {Delle Luche}, {Domingues}, {Dubath}, {Fodor}, {Fr{\'e}zouls},
  {Fries}, {Fustes}, {Fyfe}, {Gallardo}, {Gallegos}, {Gardiol}, {Gebran},
  {Gomboc}, {G{\'o}mez}, {Grux}, {Gueguen}, {Heyrovsky}, {Hoar}, {Iannicola},
  {Isasi Parache}, {Janotto}, {Joliet}, {Jonckheere}, {Keil}, {Kim},
  {Klagyivik}, {Klar}, {Knude}, {Kochukhov}, {Kolka}, {Kos}, {Kutka}, {Lainey},
  {LeBouquin}, {Liu}, {Loreggia}, {Makarov}, {Marseille}, {Martayan},
  {Martinez-Rubi}, {Massart}, {Meynadier}, {Mignot}, {Munari}, {Nguyen},
  {Nordlander}, {Ocvirk}, {O'Flaherty}, {Olias Sanz}, {Ortiz}, {Osorio},
  {Oszkiewicz}, {Ouzounis}, {Palmer}, {Park}, {Pasquato}, {Peltzer}, {Peralta},
  {P{\'e}turaud}, {Pieniluoma}, {Pigozzi}, {Poels}, {Prat}, {Prod'homme},
  {Raison}, {Rebordao}, {Risquez}, {Rocca-Volmerange}, {Rosen}, {Ruiz-Fuertes},
  {Russo}, {Sembay}, {Serraller Vizcaino}, {Short}, {Siebert}, {Silva},
  {Sinachopoulos}, {Slezak}, {Soffel}, {Sosnowska}, {Strai{\v{z}}ys}, {ter
  Linden}, {Terrell}, {Theil}, {Tiede}, {Troisi}, {Tsalmantza}, {Tur},
  {Vaccari}, {Vachier}, {Valles}, {Van Hamme}, {Veltz}, {Virtanen}, {Wallut},
  {Wichmann}, {Wilkinson}, {Ziaeepour}, \& {Zschocke}}]{GaiaMission2016}
{Gaia Collaboration}, {Prusti}, T., {de Bruijne}, J.~H.~J., {et~al.} 2016,
  \aap, 595, A1

\bibitem[{{Gaia Collaboration} {et~al.}(2022{\natexlab{b}}){Gaia
  Collaboration}, {Vallenari, A.}, {Brown, A.G.A.}, {Prusti, T.}, \& {et
  al.}}]{GaiaDR3}
{Gaia Collaboration}, {Vallenari, A.}, {Brown, A.G.A.}, {Prusti, T.}, \& {et
  al.} 2022{\natexlab{b}}, A\&A

\bibitem[{{G{\'o}rski} {et~al.}(2005){G{\'o}rski}, {Hivon}, {Banday},
  {Wandelt}, {Hansen}, {Reinecke}, \& {Bartelmann}}]{healpy}
{G{\'o}rski}, K.~M., {Hivon}, E., {Banday}, A.~J., {et~al.} 2005, \apj, 622,
  759

\bibitem[{Harris {et~al.}(2020)Harris, Millman, van~der Walt, Gommers,
  Virtanen, Cournapeau, Wieser, Taylor, Berg, Smith, Kern, Picus, Hoyer, van
  Kerkwijk, Brett, Haldane, del R{'{\i}}o, Wiebe, Peterson,
  G{'{e}}rard-Marchant, Sheppard, Reddy, Weckesser, Abbasi, Gohlke, \&
  Oliphant}]{numpy}
Harris, C.~R., Millman, K.~J., van~der Walt, S.~J., {et~al.} 2020, Nature, 585,
  357

\bibitem[{{Hunt} {et~al.}(2022){Hunt}, {Price-Whelan}, {Johnston}, \&
  {Darragh-Ford}}]{Hunt2022}
{Hunt}, J. A.~S., {Price-Whelan}, A.~M., {Johnston}, K.~V., \& {Darragh-Ford},
  E. 2022, \mnras, 516, L7

\bibitem[{{Hunt} {et~al.}(2021){Hunt}, {Stelea}, {Johnston}, {Gandhi},
  {Laporte}, \& {B{\'e}dorf}}]{2021MNRAS.508.1459H}
{Hunt}, J. A.~S., {Stelea}, I.~A., {Johnston}, K.~V., {et~al.} 2021, \mnras,
  508, 1459

\bibitem[{Hunter(2007)}]{matplotlib}
Hunter, J.~D. 2007, Computing In Science \& Engineering, 9, 90

\bibitem[{{Kingma} \& {Ba}(2014)}]{adamopt}
{Kingma}, D.~P. \& {Ba}, J. 2014, arXiv e-prints, arXiv:1412.6980

\bibitem[{{Kordopatis} {et~al.}(2022){Kordopatis}, {Schultheis}, {McMillan},
  {Palicio}, {de Laverny}, {Recio-Blanco}, {Creevey}, {{\'A}lvarez}, {Andrae},
  {Poggio}, {Spitoni}, {Contursi}, {Zhao}, {Oreshina-Slezak}, {Ordenovic}, \&
  {Bijaoui}}]{Kordopatis2022}
{Kordopatis}, G., {Schultheis}, M., {McMillan}, P.~J., {et~al.} 2022, arXiv
  e-prints, arXiv:2206.07937

\bibitem[{{Kumar} {et~al.}(2022){Kumar}, {Ghosh}, {Kataria}, {Das}, \&
  {Debattista}}]{Kumar2022}
{Kumar}, A., {Ghosh}, S., {Kataria}, S.~K., {Das}, M., \& {Debattista}, V.~P.
  2022, \mnras, 516, 1114

\bibitem[{{Laporte} {et~al.}(2019){Laporte}, {Minchev}, {Johnston}, \&
  {G{\'o}mez}}]{laporte2019}
{Laporte}, C. F.~P., {Minchev}, I., {Johnston}, K.~V., \& {G{\'o}mez}, F.~A.
  2019, \mnras, 485, 3134

\bibitem[{{Li} \& {Widrow}(2021)}]{li2021}
{Li}, H. \& {Widrow}, L.~M. 2021, \mnras, 503, 1586

\bibitem[{{Martinez-Medina} {et~al.}(2022){Martinez-Medina},
  {P{\'e}rez-Villegas}, \& {Peimbert}}]{2022MNRAS.512.1574M}
{Martinez-Medina}, L., {P{\'e}rez-Villegas}, A., \& {Peimbert}, A. 2022,
  \mnras, 512, 1574

\bibitem[{{Mathur}(1990)}]{mathur1990}
{Mathur}, S.~D. 1990, \mnras, 243, 529

\bibitem[{McMillan(2016)}]{mcmillan2016mass}
McMillan, P.~J. 2016, Monthly Notices of the Royal Astronomical Society,
  stw2759

\bibitem[{{Monari} {et~al.}(2015){Monari}, {Famaey}, \& {Siebert}}]{monari2015}
{Monari}, G., {Famaey}, B., \& {Siebert}, A. 2015, \mnras, 452, 747

\bibitem[{{Monari} {et~al.}(2016{\natexlab{a}}){Monari}, {Famaey}, \&
  {Siebert}}]{monari2016a}
{Monari}, G., {Famaey}, B., \& {Siebert}, A. 2016{\natexlab{a}}, \mnras, 457,
  2569

\bibitem[{{Monari} {et~al.}(2016{\natexlab{b}}){Monari}, {Famaey}, {Siebert},
  {Grand}, {Kawata}, \& {Boily}}]{monari2016b}
{Monari}, G., {Famaey}, B., {Siebert}, A., {et~al.} 2016{\natexlab{b}}, \mnras,
  461, 3835

\bibitem[{{Nelson} \& {Widrow}(2022)}]{nelson2022}
{Nelson}, P. \& {Widrow}, L.~M. 2022, \mnras, 516, 5429

\bibitem[{{Oort}(1932)}]{oort1932}
{Oort}, J.~H. 1932, \bain, 6, 249

\bibitem[{{Poggio} {et~al.}(2020){Poggio}, {Drimmel}, {Andrae}, {Bailer-Jones},
  {Fouesneau}, {Lattanzi}, {Smart}, \& {Spagna}}]{Poggio2020}
{Poggio}, E., {Drimmel}, R., {Andrae}, R., {et~al.} 2020, Nature Astronomy, 4,
  590

\bibitem[{Rasmussen \& Williams(2005)}]{10.5555/1162254}
Rasmussen, C.~E. \& Williams, C. K.~I. 2005, Gaussian Processes for Machine
  Learning (Adaptive Computation and Machine Learning) (The MIT Press)

\bibitem[{{Reid} {et~al.}(2019){Reid}, {Menten}, {Brunthaler}, {Zheng}, {Dame},
  {Xu}, {Li}, {Sakai}, {Wu}, {Immer}, {Zhang}, {Sanna}, {Moscadelli}, {Rygl},
  {Bartkiewicz}, {Hu}, {Quiroga-Nu{\~n}ez}, \& {van Langevelde}}]{Reid2019}
{Reid}, M.~J., {Menten}, K.~M., {Brunthaler}, A., {et~al.} 2019, \apj, 885, 131

\bibitem[{{Reid} {et~al.}(2014){Reid}, {Menten}, {Brunthaler}, {Zheng}, {Dame},
  {Xu}, {Wu}, {Zhang}, {Sanna}, {Sato}, {Hachisuka}, {Choi}, {Immer},
  {Moscadelli}, {Rygl}, \& {Bartkiewicz}}]{Reid2014}
{Reid}, M.~J., {Menten}, K.~M., {Brunthaler}, A., {et~al.} 2014, \apj, 783, 130

\bibitem[{{Rybizki} {et~al.}(2021){Rybizki}, {Rix}, {Demleitner},
  {Bailer-Jones}, \& {Cooper}}]{rybizki2021}
{Rybizki}, J., {Rix}, H.-W., {Demleitner}, M., {Bailer-Jones}, C. A.~L., \&
  {Cooper}, W.~J. 2021, \mnras, 500, 397

\bibitem[{{Sch{\"o}nrich} \& {Dehnen}(2018)}]{schonrich2018}
{Sch{\"o}nrich}, R. \& {Dehnen}, W. 2018, \mnras, 478, 3809

\bibitem[{{Sellwood}(2013)}]{sellwood2013}
{Sellwood}, J.~A. 2013, in Planets, Stars and Stellar Systems. Volume 5:
  Galactic Structure and Stellar Populations, ed. T.~D. {Oswalt} \&
  G.~{Gilmore}, Vol.~5, 923

\bibitem[{Titsias(2009)}]{Titsias2009}
Titsias, M. 2009, Journal of Machine Learning Research - Proceedings Track, 5,
  567

\bibitem[{{Weinberg}(1991)}]{weinberg1991}
{Weinberg}, M.~D. 1991, \apj, 373, 391

\bibitem[{{Widmark} {et~al.}(2022{\natexlab{a}}){Widmark}, {Hunt}, {Laporte},
  \& {Monari}}]{widmark2022}
{Widmark}, A., {Hunt}, J.~A.~S., {Laporte}, C.~F.~P., \& {Monari}, G.
  2022{\natexlab{a}}, \aap, 663, A16

\bibitem[{{Widmark} {et~al.}(2021){Widmark}, {Laporte}, {de Salas}, \&
  {Monari}}]{2021A&A...653A..86W}
{Widmark}, A., {Laporte}, C.~F.~P., {de Salas}, P.~F., \& {Monari}, G. 2021,
  \aap, 653, A86

\bibitem[{{Widmark} {et~al.}(2022{\natexlab{b}}){Widmark}, {Laporte}, \&
  {Monari}}]{widmark2021_spiralsIII}
{Widmark}, A., {Laporte}, C.~F.~P., \& {Monari}, G. 2022{\natexlab{b}}, \aap,
  663, A15

\bibitem[{{Widrow} \& {Bonner}(2015)}]{widrow2015}
{Widrow}, L.~M. \& {Bonner}, G. 2015, \mnras, 450, 266

\bibitem[{{Widrow} {et~al.}(2012){Widrow}, {Gardner}, {Yanny}, {Dodelson}, \&
  {Chen}}]{widrow2012}
{Widrow}, L.~M., {Gardner}, S., {Yanny}, B., {Dodelson}, S., \& {Chen}, H.-Y.
  2012, \apjl, 750, L41

\bibitem[{{Williams} {et~al.}(2013){Williams}, {Steinmetz}, {Binney},
  {Siebert}, {Enke}, {Famaey}, {Minchev}, {de Jong}, {Boeche}, {Freeman},
  {Bienaym{\'e}}, {Bland-Hawthorn}, {Gibson}, {Gilmore}, {Grebel}, {Helmi},
  {Kordopatis}, {Munari}, {Navarro}, {Parker}, {Reid}, {Seabroke}, {Sharma},
  {Siviero}, {Watson}, {Wyse}, \& {Zwitter}}]{williams2013}
{Williams}, M.~E.~K., {Steinmetz}, M., {Binney}, J., {et~al.} 2013, \mnras,
  436, 101

\bibitem[{{Xu} {et~al.}(2013){Xu}, {Li}, {Reid}, {Menten}, {Zheng},
  {Brunthaler}, {Moscadelli}, {Dame}, \& {Zhang}}]{Xu2013}
{Xu}, Y., {Li}, J.~J., {Reid}, M.~J., {et~al.} 2013, \apj, 769, 15

\bibitem[{{Xu} {et~al.}(2015){Xu}, {Newberg}, {Carlin}, {Liu}, {Deng}, {Li},
  {Sch{\"o}nrich}, \& {Yanny}}]{xu2015}
{Xu}, Y., {Newberg}, H.~J., {Carlin}, J.~L., {et~al.} 2015, \apj, 801, 105

\bibitem[{{Yanny} \& {Gardner}(2013)}]{yanny2013}
{Yanny}, B. \& {Gardner}, S. 2013, \apj, 777, 91

\end{thebibliography}

\begin{appendix} 

\section{Stellar ages}\label{app:ages}

\begin{figure}
	\centering
	\includegraphics[width=1\columnwidth]{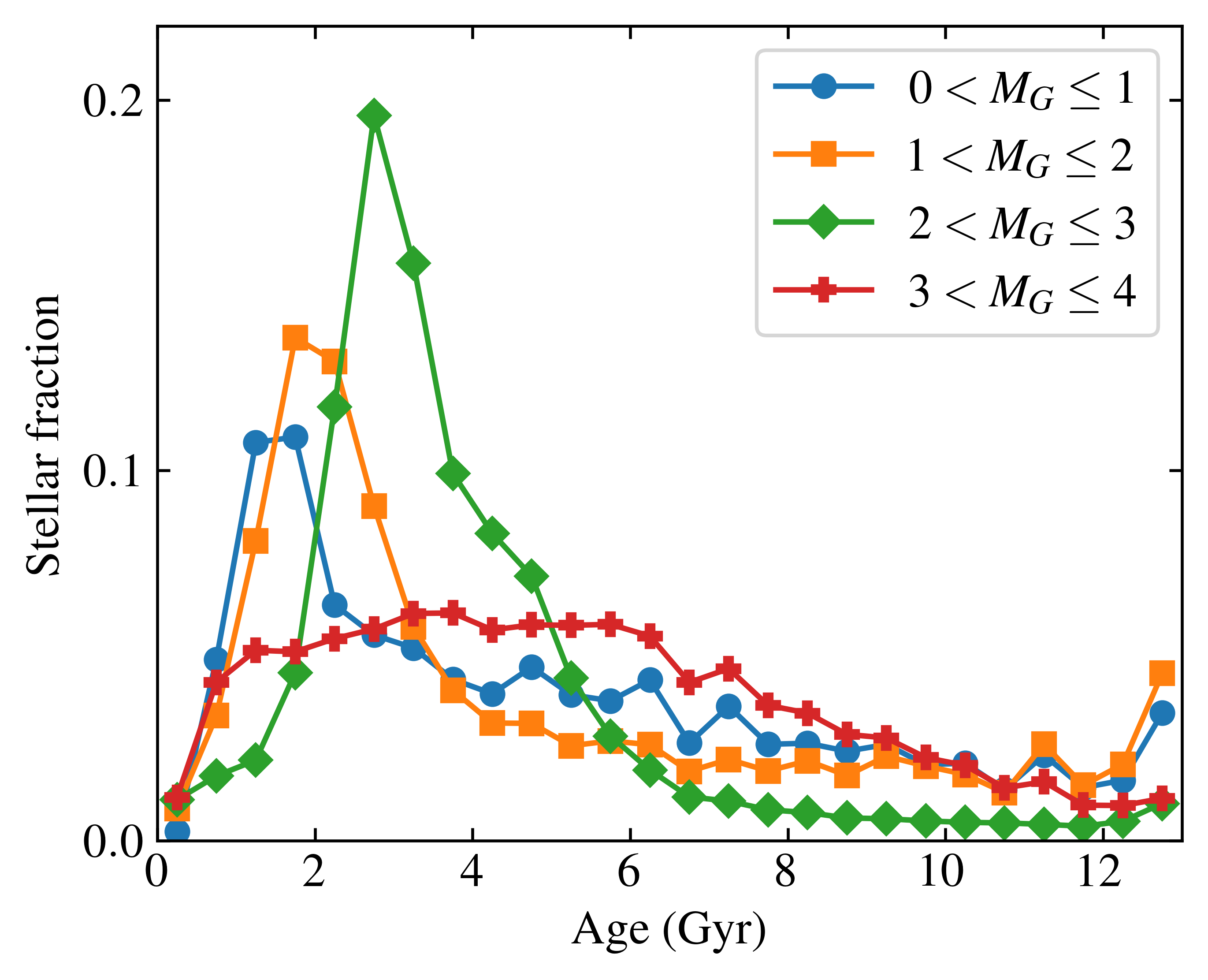}
    \caption{
    Stellar age distribution for the four absolute magnitude cuts of our data samples based on stars within 400~pc of the Sun. The markers denote the mid-point of age bins with a width of 500~Myr, in range 0--13~Gyr. The vertical axis shows the relative fraction of stars in each bin, normalised such that the area under curve is unity in these units.
    }
    \label{fig:ages}
\end{figure}

In Fig.~\ref{fig:ages}, we show the age distribution for stars within 400~pc of the Sun, which was the same as the distance range used in Fig.~\ref{fig:CMD}. The ages were taken from the catalogue by \cite{Kordopatis2022}, which were inferred using spectroscopically derived atmospheric parameters in combination with 2MASS and \emph{Gaia} photometry. The numbers of stars used for the four distributions were $19k, 13k, 74k,$ and  $281k$. Thus, this age distribution comes from a small subset of stars observed with \emph{Gaia}, but nonetheless gives an idea about the age range of our respective data samples.

\clearpage

\onecolumn

\begin{table*}[ht!]
    \centering
    \begin{tabular}{c||c|c|c|c}
                    & $0 < M_G \leq 1$ & $1 < M_G \leq 2$ & $2 < M_G \leq 3$ & $3 < M_G \leq 4$              \\
        \hline
        \hline
        $a_i$       & $\{5.39, 2.56, 4.06\}$     & $\{10.61, 5.01, 1.42\}$     & $\{10.08, 21.09, 7.72\}$    & $\{17.21, 53.84, 18.72\}$       \\
        $L_i$       & $\{1524, 1526, 4563\}~\pc$ & $\{1436, 5158, 5917\}~\pc$ & $\{990, 2565, 16338\}~\pc$ & $\{821, 2218, 6576\}~\pc$    \\
        $h_i$       & $\{129, 130, 465\}~\pc$    & $\{102, 255, 651\}~\pc$    & $\{105, 205, 497\}~\pc$    & $\{115, 296, 667\}~\pc$       \\
        $Z_\odot$   & 7.47 pc                    & 6.64 pc                    & 12.7 pc                    & 18.6 pc                       \\
    \end{tabular}
    \caption{Inferred parameters for $f_\text{symm.}$, for our four data samples.}
    \label{tab:symm_fit_res}
\end{table*}

\begin{multicols}{2}

\section{Symmetric analytic function fitted parameters}\label{app:symm_fit_res}

The fitted free parameters of $f_\text{symm.}$ can be found in Table~\ref{tab:symm_fit_res}. We also performed fits with a smaller or larger number of disk components. Using more than three disk components did not produce noticeably better fits. Using only two disk components did in fact produce some artefacts, because in that case $f_\text{symm.}$ could not replicate the heavy tails towards high $|Z|$. An exception is the very brightest data sample, for which two disk components have practically identical scale length and scale height values, such that only two disk component would suffice. The scale lengths and heights of the three disk components are increasing in unison, for all four data samples. The Sun's height with respect to the disk mid-plane is found to be roughly 11~pc, with variations of a few parsec between the data samples.

As an additional test, we modified the functional form of $f_\text{symm.}$ to read
\begin{equation}
\begin{split}
    & f_\text{symm.}(X,Y,Z \, | \, a_i,L_i,H_i,Z_\odot,\alpha,\beta) = \\ 
    & \sum_{i=1}^{3}
    a_i \, \exp \left(-\frac{R-R_\odot}{L_i} \right) \, \text{sech}^2
    \left( \frac{Z+Z_\odot+\alpha X+\beta Y}{H_i} \right).
\end{split}
\end{equation}
This differs from Eq.~\ref{eq:f_symm} in that we have added the $\alpha$ and $\beta$ parameters, which correspond to a potential inclination of the disk mid-plane, different from the plane defined by $b=0~\deg$ in the \emph{Gaia} catalogue. However, our results for the plane inclination are minimal; we infer ($\alpha$, $\beta$) values of (0.00025, -0.00591), (-0.00001, -0.00612), (0.00176, -0.00725), and (0.00407, -0.00634) for our four data samples. These values mean that at a 2~kpc distance from the Sun, the disk mid-plane varies on the scale of roughly 10~pc as compared to the plane defined by $b=0~\deg$. We see slight evidence for a misalignment between these two planes, but this result could very well be affected by systematic errors. Either way, this misalignment is not strong enough to alter the general conclusions of this work.

\newpage

\end{multicols}

\clearpage

\twocolumn

\section{Supplementary figures}\label{sec:other_plots}

\begin{figure*}[h]
	\centering
	\includegraphics[width=0.6\textwidth]{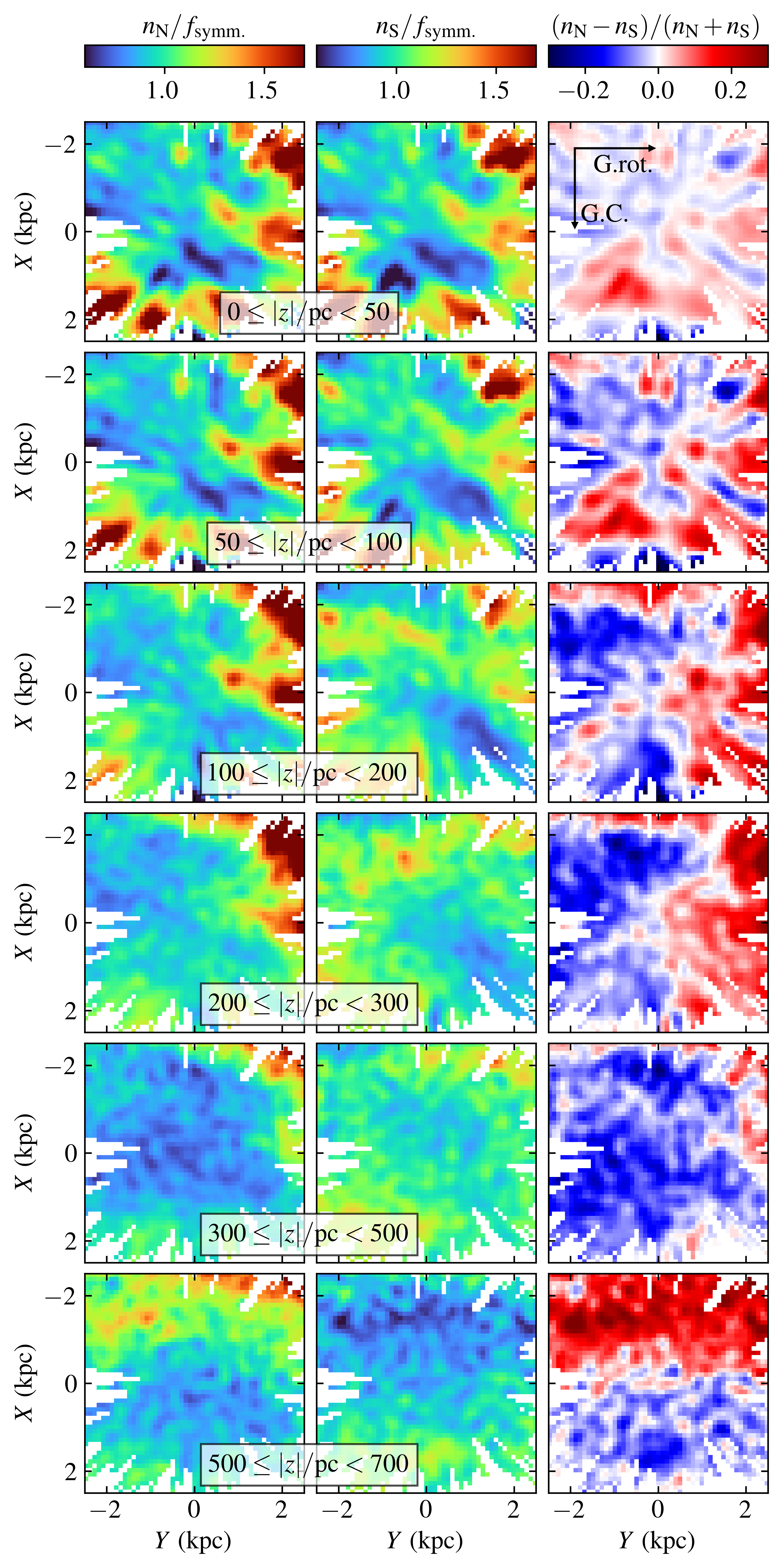}
    \caption{Same as Fig.~\ref{fig:disk_residuals_2<G<3}, but for the stellar sample with $0 < M_G \leq 1$.}
    \label{fig:disk_residuals_0<G<1}
\end{figure*}

\begin{figure*}[h]
	\centering
	\includegraphics[width=0.6\textwidth]{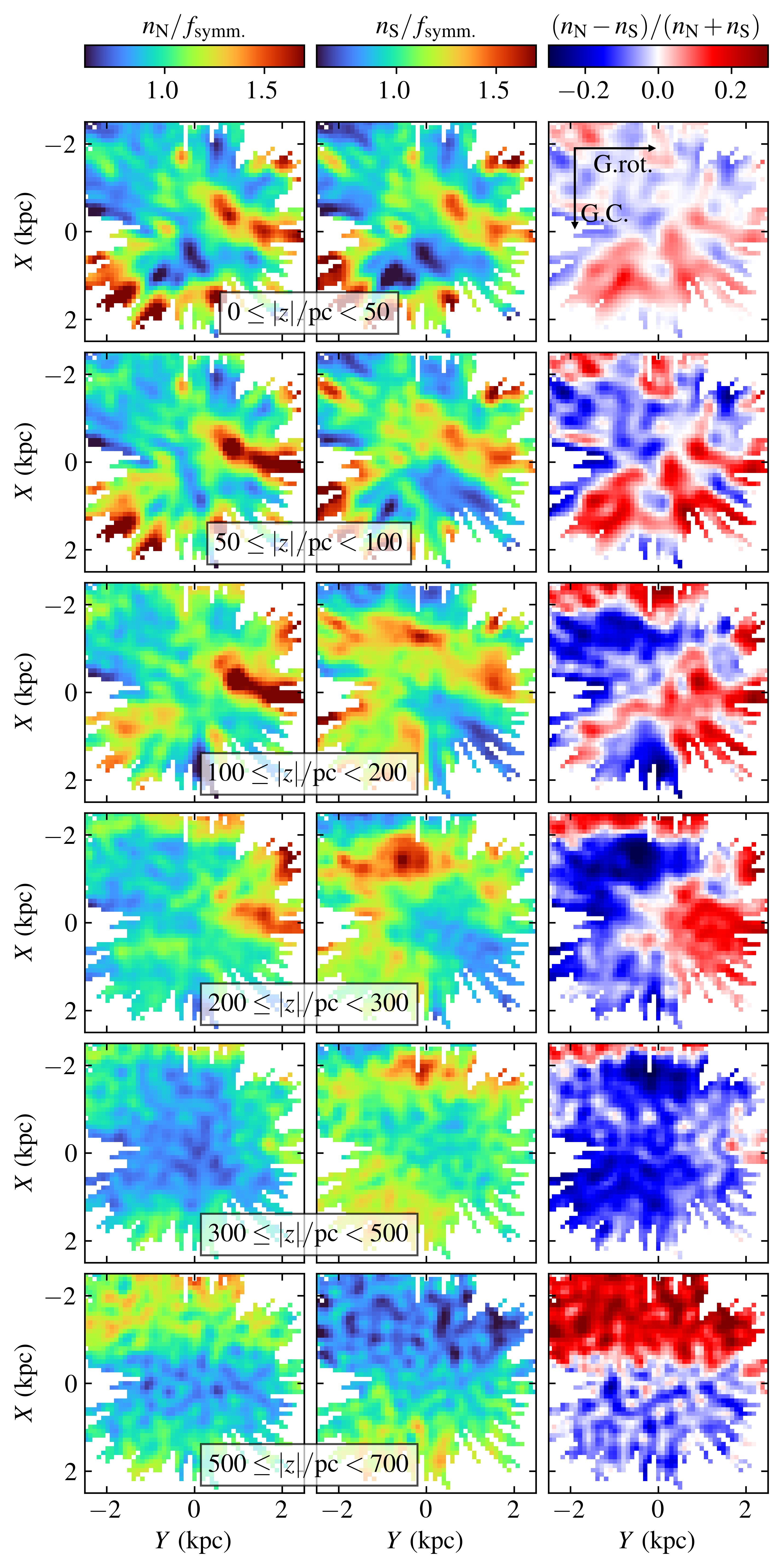}
    \caption{Same as Fig.~\ref{fig:disk_residuals_2<G<3}, but for the stellar sample with $1 < M_G \leq 2$.}
    \label{fig:disk_residuals_1<G<2}
\end{figure*}

\begin{figure*}[h]
	\centering
	\includegraphics[width=0.6\textwidth]{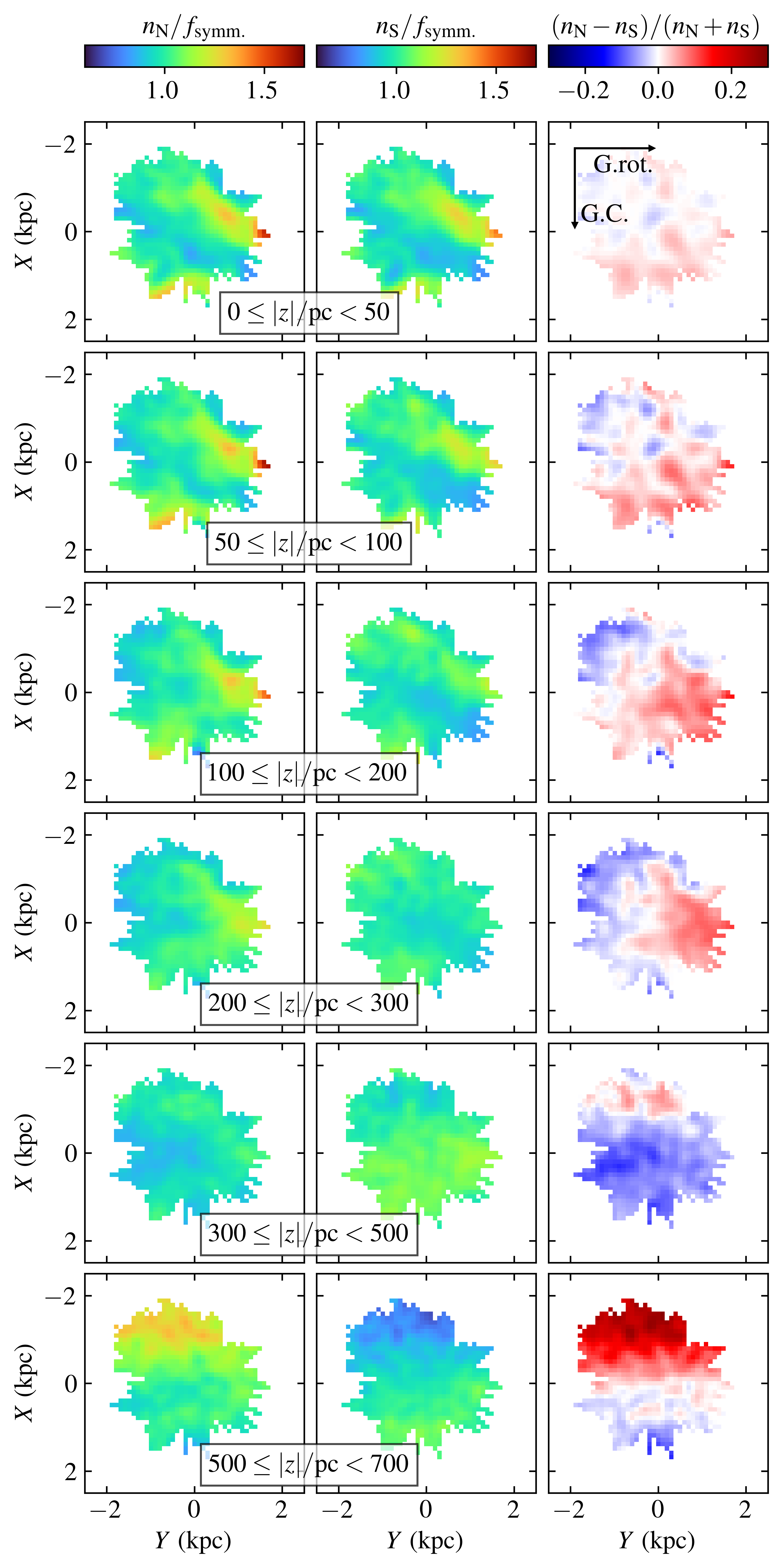}
    \caption{Same as Fig.~\ref{fig:disk_residuals_2<G<3}, but for the stellar sample with $3 < M_G \leq 4$.}
    \label{fig:disk_residuals_3<G<4}
\end{figure*}

\begin{figure}[h]
	\centering
	\includegraphics[width=1\columnwidth]{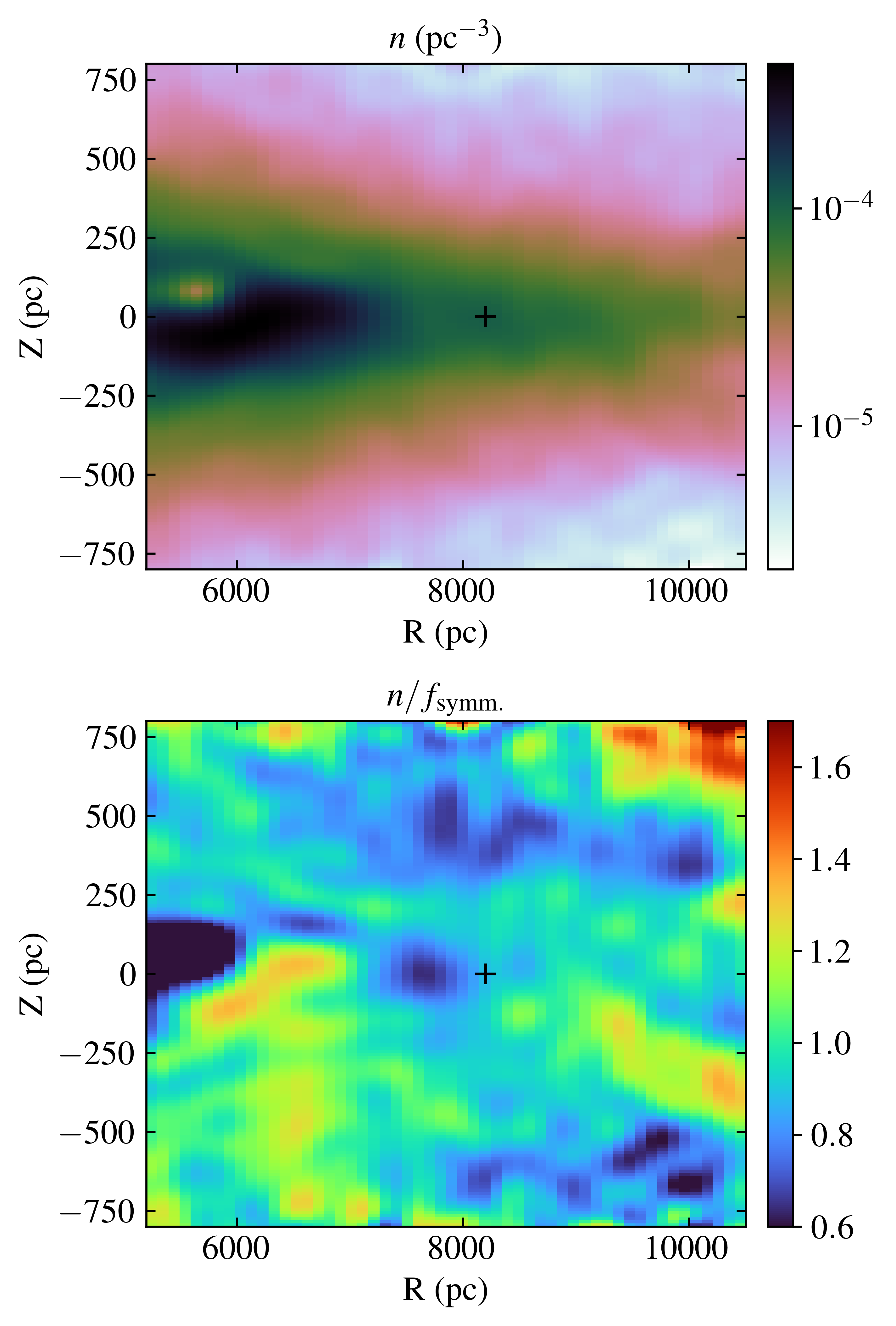}
    \caption{Like Fig.~\ref{fig:Rz_residuals_2<G<3}, but for the stellar sample with $0 < M_G \leq 1$.}
    \label{fig:Rz_residuals_0<G<1}
\end{figure}

\begin{figure}[h]
	\centering
	\includegraphics[width=1\columnwidth]{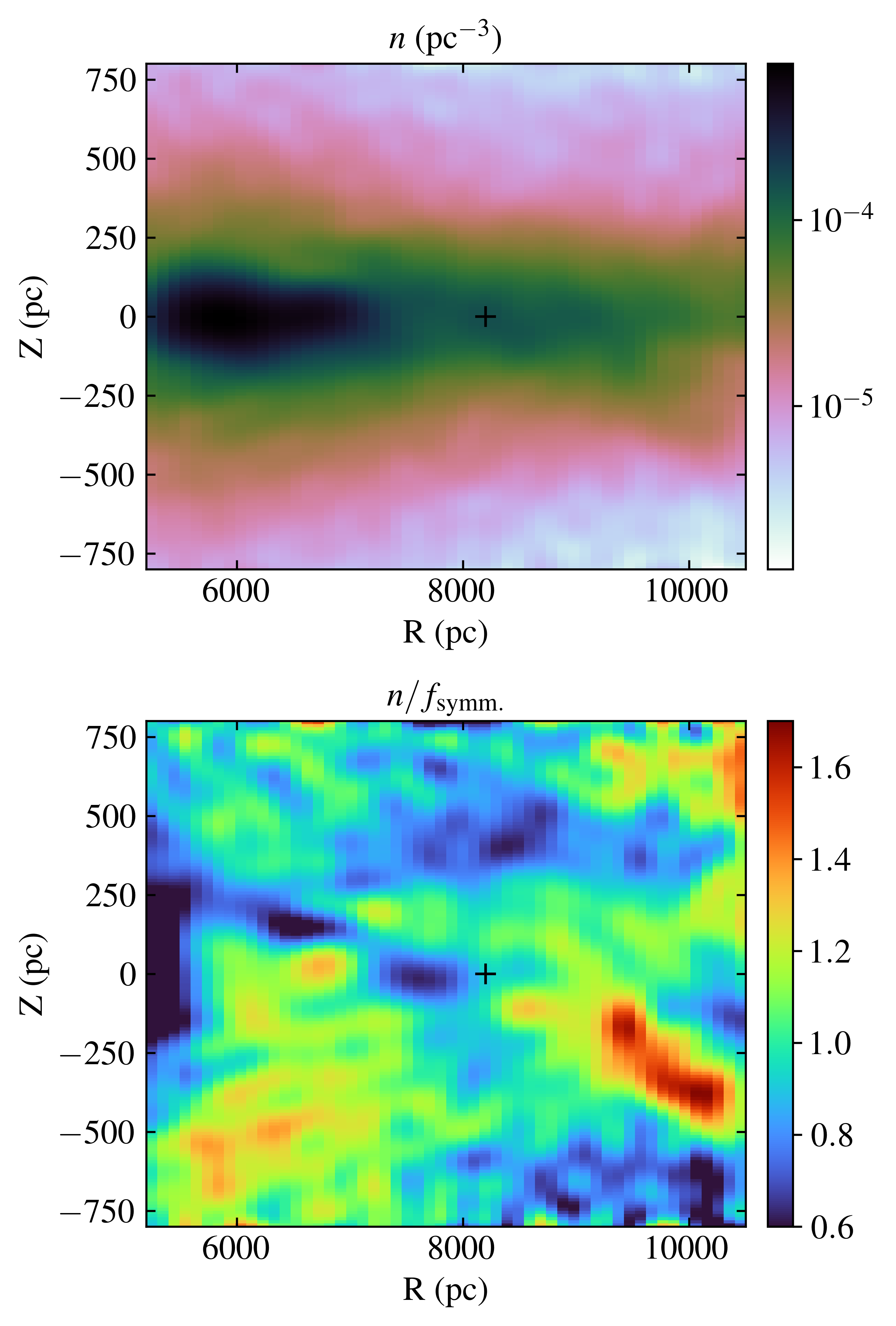}
    \caption{Like Fig.~\ref{fig:Rz_residuals_2<G<3}, but for the stellar sample with $1 < M_G \leq 2$.}
    \label{fig:Rz_residuals_1<G<2}
\end{figure}

\begin{figure}[h]
	\centering
	\includegraphics[width=1\columnwidth]{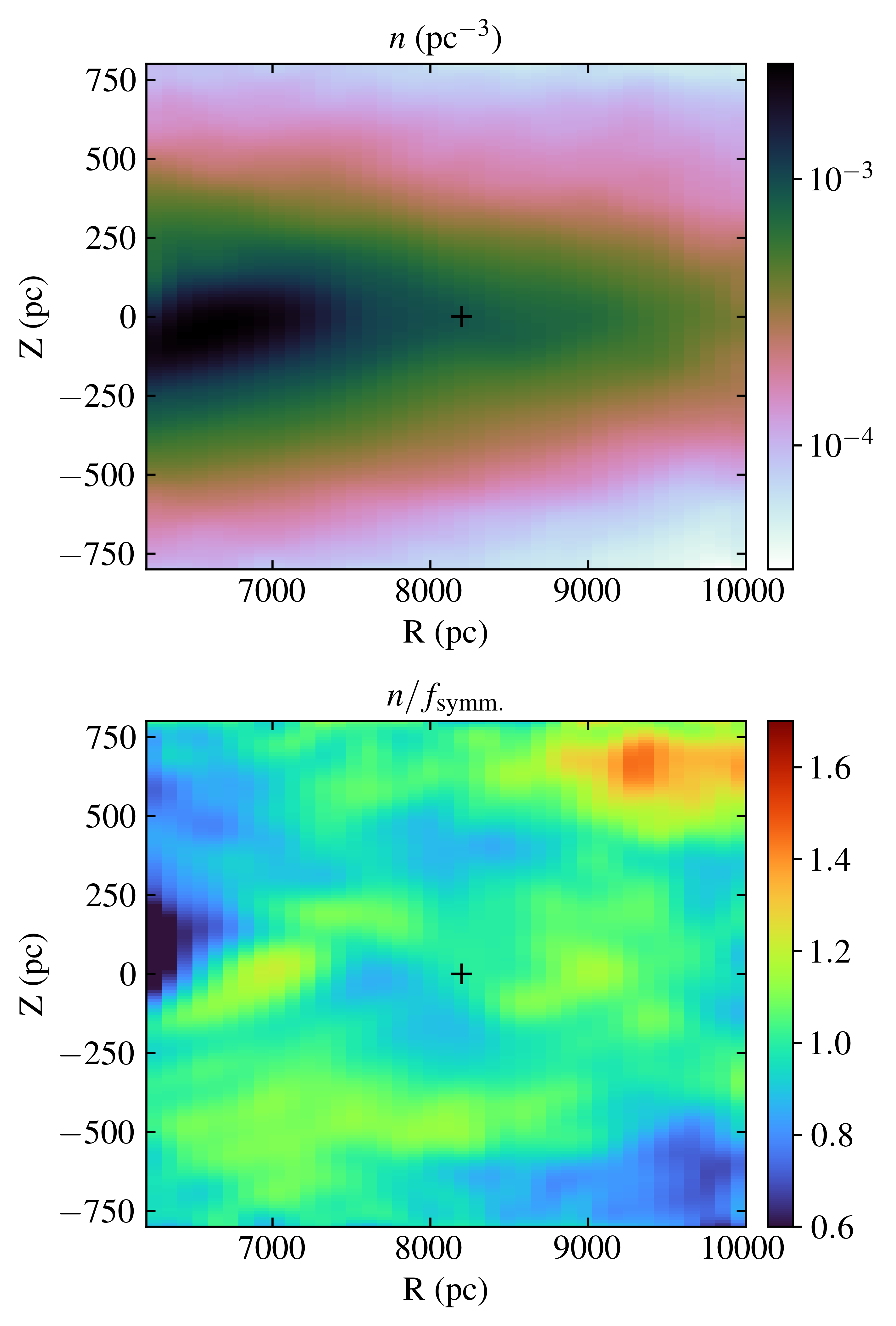}
    \caption{Like Fig.~\ref{fig:Rz_residuals_2<G<3}, but for the stellar sample with $3 < M_G \leq 4$.}
    \label{fig:Rz_residuals_3<G<4}
\end{figure}

\begin{figure*}[h]
	\centering
	\includegraphics[width=0.96\textwidth]{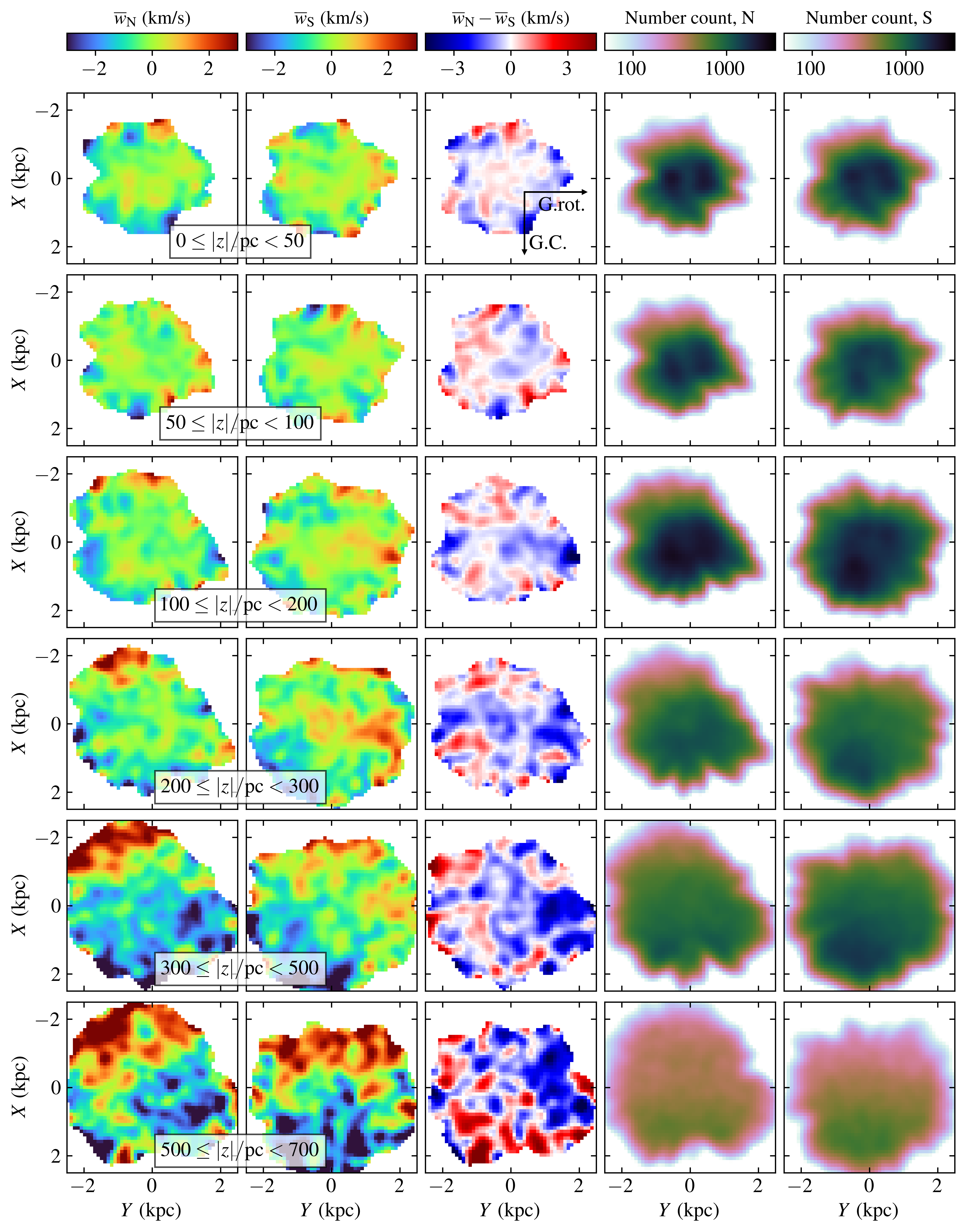}
    \caption{Same as Fig.~\ref{fig:disk_W_0<G<1}, but for the stellar sample with $1 < M_G \leq 2$.}
    \label{fig:disk_W_1<G<2}
\end{figure*}

\begin{figure*}[h]
	\centering
	\includegraphics[width=0.96\textwidth]{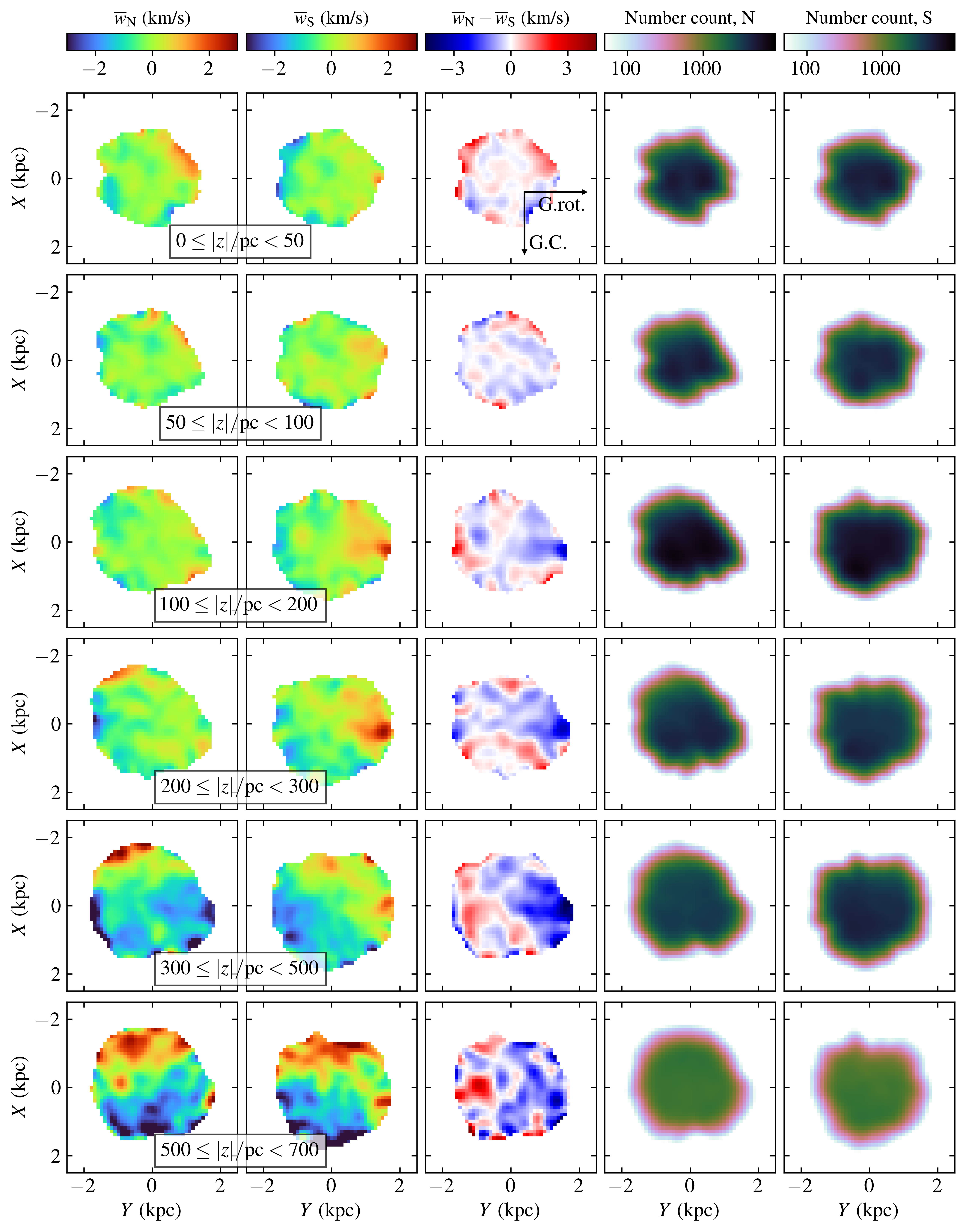}
    \caption{Same as Fig.~\ref{fig:disk_W_0<G<1}, but for the stellar sample with $2 < M_G \leq 3$.}
    \label{fig:disk_W_2<G<3}
\end{figure*}

In Figs.~\ref{fig:disk_residuals_0<G<1}--\ref{fig:disk_W_2<G<3}, we show plots corresponding to Figs.~\ref{fig:disk_residuals_2<G<3}, \ref{fig:Rz_residuals_2<G<3}, and \ref{fig:disk_W_0<G<1} in the main text, but for our other data samples (although the velocity plot for our dimmest data sample is excluded due to covering such a small spatial volume). For brighter data samples, in the $(X,Y)$-plane projections of Figs.~\ref{fig:disk_residuals_0<G<1} and \ref{fig:disk_residuals_1<G<2}, as well as the $(R,z)$-plane projections of Figs.~\ref{fig:Rz_residuals_0<G<1} and \ref{fig:Rz_residuals_1<G<2}, the distant regions ($\gtrsim 2~\kpc$) seem to suffer from strong systematic errors, especially close to the disk mid-plane.


\clearpage

\section{Spiral angle plots}
\label{app:spiral_plots}

\begin{figure}[h]
	\centering
	\includegraphics[width=1\columnwidth]{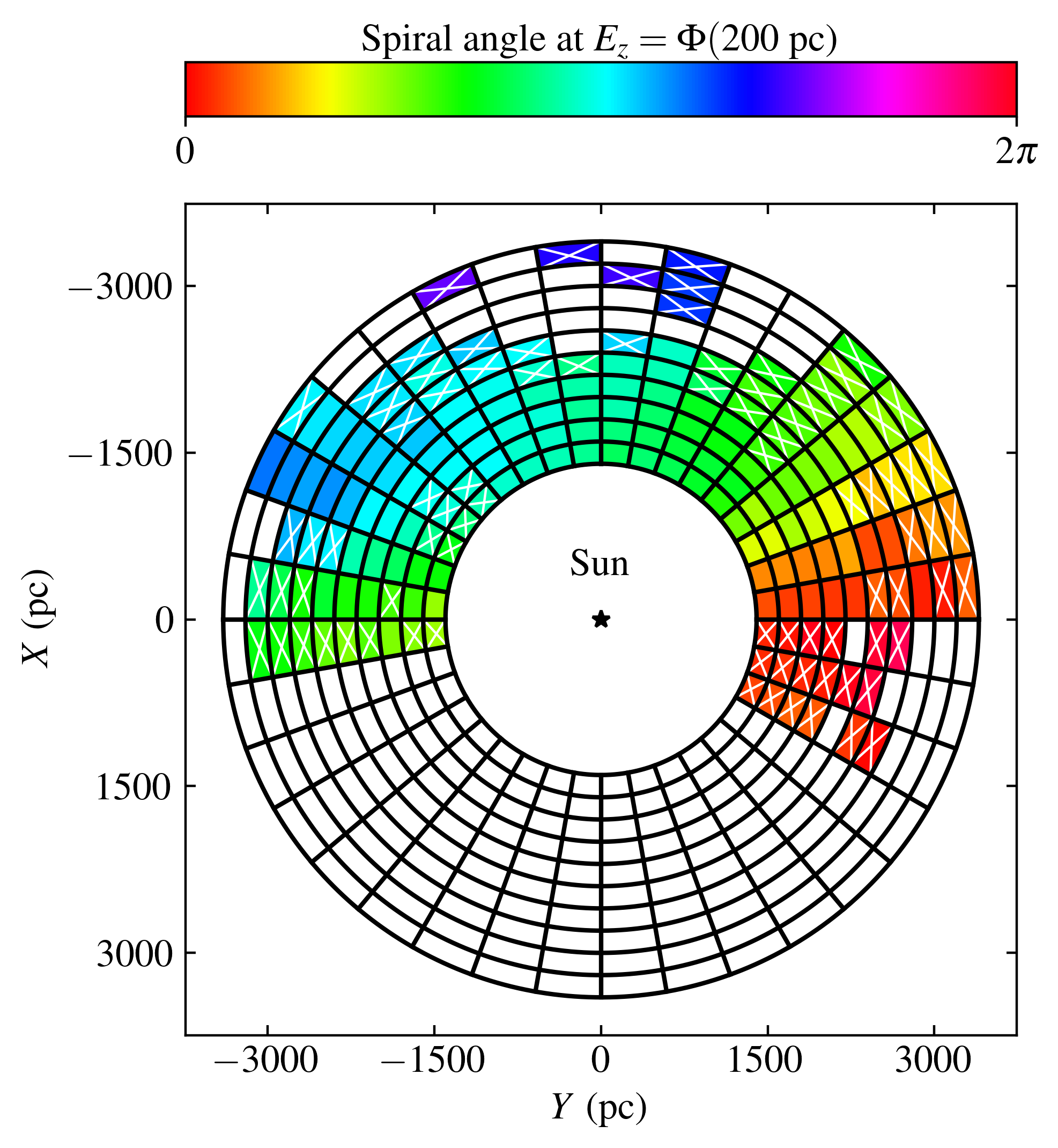}
    \caption{Angle of the phase-space spiral at the iso-energy contour $E_z=\Phi(200~\pc)$ in the $(z,w)$-plane, as inferred in \citet{widmark2021_spiralsIII}. The colour bar is cyclical. If the angle is close to 0 (or $2\pi$, equivalently), then the spiral perturbation corresponds to an over-density at the phase-space coordinates $(Z,W)=(200~\pc, 0~\kmsec)$; conversely, if the angle is close to $\pi$, then the spiral corresponds to an under-density at $(Z,W)=(200~\pc, 0~\kmsec)$. Area cells that are crossed over in white are marked as less trustworthy; we refer to \citet{widmark2021_spiralsIII} for details.
    }
    \label{fig:spiral_angle_200pc}
\end{figure}

\begin{figure}[h]
	\centering
	\includegraphics[width=1\columnwidth]{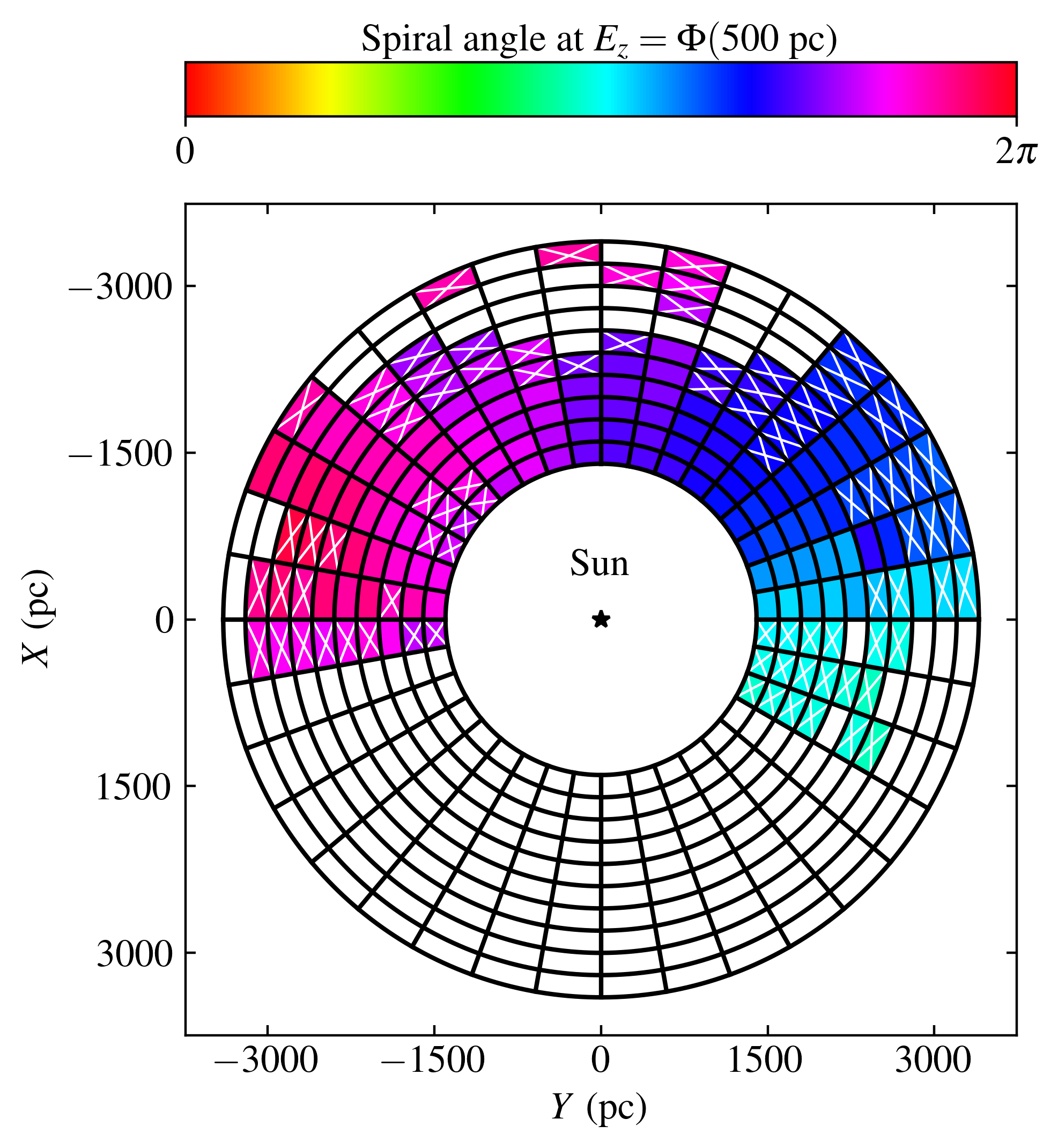}
    \caption{Same as Fig.~\ref{fig:spiral_angle_200pc} but for the iso-energy contour $E_z=\Phi(500~\pc)$.}
    \label{fig:spiral_angle_500pc}
\end{figure}

In Figs.~\ref{fig:spiral_angle_200pc} and \ref{fig:spiral_angle_500pc}, we show how the spiral angle varies in the $(X,Y)$-plane. The spiral angle, more specifically, is given by the location of the phase-space spiral over-density in the $(z,w)$-plane along the iso-contour of vertical energy. In the two plots, this vertical energy is fixed to either $E_z = \Phi(200~\pc)$ or $E_z = \Phi(500~\pc)$. These results come from directly from \citet{widmark2021_spiralsIII}, although this figure was not included in that article; we refer to that article for further details.

In Fig.~\ref{fig:spiral_angle_200pc} and $E_z = \Phi(200~\pc)$, an angle close to 0 (or $2\pi$, equivalently), means that the spiral perturbation corresponds to an over-density, when projected onto the $Z$-axis, at the height $Z\simeq 200~\pc$. Because the spiral is single-armed and asymmetric, that also means that there is an under-density in $n(Z)$ at $Z\simeq -200~\pc$. This agrees well with the north-south asymmetry seen in the third and fourth rows of Figs.~\ref{fig:disk_residuals_2<G<3}, \ref{fig:disk_residuals_0<G<1}--\ref{fig:disk_residuals_3<G<4}, which has a corresponding morphology in the $(X,Y)$-plane, for example in terms of an over-density region in the direction of positive $Y$. The spiral's projected perturbation at greater heights is less clear. The projected spiral density perturbation at $Z\simeq 500~\pc$, as seen in Fig.~\ref{fig:spiral_angle_500pc}, does not have a clear counterpart in the fourth or fifth rows of of Fig.~\ref{fig:disk_residuals_2<G<3}, indicating a superposition of other significant asymmetries at these greater heights.

It is also evident from these figures that the spiral angle varies significantly with the azimuth, over scales of a few kiloparsecs. As such, the morphology of the phase-space spiral does not match the two main perturbation features that we identify in this work (the small scale breathing mode associated with the Local Spiral Arm, and the large-scale bending mode), whose properties vary on either much smaller or much larger scales in the $(X,Y)$-plane.

\end{appendix}

\end{document}